\documentclass[a4paper,fleqn]{cas-dc}

\usepackage[numbers]{natbib}
\usepackage{bm}
\usepackage{algorithm}
\usepackage{algpseudocode}
\usepackage{subcaption}
\usepackage{url}
\usepackage{tikz}
\usetikzlibrary{arrows.meta,positioning,shapes.geometric}

\ExplSyntaxOn
\cs_set:Npn \__reset_fig:
{
  \tl_set:Nn \l_fig_pos_tl { !htbp }
  \tl_set:Nx \l_fig_cols_tl { 1 }
  \tl_set:Nn \l_fig_align_tl { \raggedleft }
  \skip_set:Nn \l_fig_abovecap_skip { 6pt }
  \skip_set:Nn \l_fig_belowcap_skip { 6pt }
  \skip_set:Nn \l_fig_abovefig_skip { 6pt }
  \skip_set:Nn \l_fig_belowfig_skip { 6pt }
}
\cs_set:Npn \__reset_tbl:
{
  \tl_set:Nn \l_tbl_pos_tl { !htbp }
  \tl_set:Nx \l_tbl_cols_tl { 1 }
  \tl_set:Nn \l_tbl_align_tl { \centering }
  \skip_set:Nn \l_tbl_abovecap_skip { 6pt }
  \skip_set:Nn \l_tbl_belowcap_skip { 0pt }
  \skip_set:Nn \l_tbl_abovetbl_skip { 6pt }
  \skip_set:Nn \l_tbl_belowtbl_skip { 6pt }
}
\ExplSyntaxOff

\AtBeginEnvironment{tabular}{\normalfont}

\makeatletter
\newenvironment{breakablealgorithm}
  {%
   \begin{center}
     \refstepcounter{algorithm}%
     \hrule height.8pt depth0pt \kern2pt%
     \renewcommand{\caption}[2][\relax]{%
       {\raggedright\textbf{\ALG@name~\thealgorithm} ##2\par}%
       \ifx\relax##1\relax
         \addcontentsline{loa}{algorithm}{\protect\numberline{\thealgorithm}##2}%
       \else
         \addcontentsline{loa}{algorithm}{\protect\numberline{\thealgorithm}##1}%
       \fi
       \kern2pt\hrule\kern2pt
     }
  }{%
     \kern2pt\hrule\relax
   \end{center}
  }
\makeatother

\begin{document}
\let\WriteBookmarks\relax
\shorttitle{Visible window generation for regional SAR}
\shortauthors{L. Li et al.}

\title[mode=title]{Efficient Generation and Quality Screening of Visible Windows for Regional SAR Reconnaissance}

\author[1,2]{Linhong Li}
\credit{Conceptualization, Methodology, Software, Writing -- original draft}

\author[1,2]{Qi Feng}
\credit{Software, Validation}

\author[1,2]{Kebo Li}
\credit{Supervision, Funding acquisition}

\author[1,2]{Yangang Liang}
\cormark[1]
\ead{Liangyg@nudt.edu.cn}
\credit{Supervision, Funding acquisition, Writing -- review \& editing}

\affiliation[1]{organization={College of Aerospace Science and Engineering, National University of Defense Technology},
                city={Changsha},
                postcode={410073},
                country={China}}

\affiliation[2]{organization={State Key Laboratory of Space System Operation and Control},
                city={Changsha},
                postcode={410073},
                country={China}}

\cortext[cor1]{Corresponding author}

\fntext[fnd1]{This work was supported in part by the National Natural Science Foundation of China under Grant 12072366 and Grant U2441205, and granted by State Key Laboratory of Space System Operation and Control.}

\begin{abstract}
Regional synthetic aperture radar reconnaissance requires observation windows that satisfy geometric feasibility under side-looking constraints and deliver interpretable image quality. This paper develops an efficient framework for visible window generation and per-window signal-level quality assessment. Window construction proceeds through three stages: coarse angular bandpass screening eliminates orbit arcs without potential target intersection, a planar characteristic curve containment test on the sensor calculation plane determines the precise geometry feasible intervals, and one-dimensional boundary bisection resolves each entry and exit epoch to subsecond precision. Each geometry feasible window then undergoes a companion point target stripmap simulation that measures range and azimuth impulse response width, peak sidelobe ratio, and integrated sidelobe ratio against mission-dependent acceptance thresholds. Numerical experiments validate the three-stage generation pipeline against an independent STK reference and demonstrate that the quality screening procedure differentiates imaging performance across observation windows with measurably different geometry. The framework provides an auditable preprocessing stage that converts continuous time regional visibility into quality-qualified observation windows suitable for subsequent mission planning.
\end{abstract}


\begin{highlights}
\item Three-stage fast window generation for regional SAR under side-looking constraints.
\item Per-window signal-level quality screening via echo simulation and back projection.
\item Quality screening differentiates imaging across windows with different geometry.
\end{highlights}

\begin{keywords}
Characteristic curve window \sep per-window quality screening \sep back projection \sep synthetic aperture radar \sep visible time window
\end{keywords}

\maketitle

\section{Introduction}
\label{sec:introduction}

Time critical reconnaissance, disaster response, maritime surveillance, and persistent monitoring continue to raise the demands placed on synthetic aperture radar (SAR) acquisitions. As mission assessment shifts from the simple question of whether a target is visible to the broader question of whether interpretable imagery can be formed, the value of an imaging activity increasingly depends on the end-to-end chain from raw echo data through image formation to focused image quality. From this perspective, a geometrically visible interval is only a potential opportunity rather than an executable observation window. For side looking SAR, incidence, squint, and dwell constraints further compress the geometrically visible arcs. Geometric accessibility alone does not guarantee interpretable imagery, yet most existing preprocessing pipelines do not test whether the fully focused image from a candidate window meets prescribed quality standards. A unified framework is needed that can both generate candidate SAR observation windows efficiently and qualify them through raw data simulation, precise focusing, and per-window quality measurement before scheduling. Guided by this viewpoint, the literature reviewed below is organized along two related axes: rapid field of view and window generation, and per-window signal-level quality evaluation.

Fast visibility prediction has a long lineage in astrodynamics. Early studies established cumulative coverage and rapid in-view computation for points, ground stations, and simple fields of view~\cite{morrison1973coverage,casten1981coverage,middour1989coverage,lawton1987inview,alfano1992visibility,mai2001imaging_comm}. Later work accelerated rise and set prediction and multi-restriction prediction through analytical or adaptive techniques~\cite{sun2012apchi,han2017interpolation_visibility}, and extended these methods to agile field of regard and oblate Earth models~\cite{wang2019jatis_for,nugnes2019oblate_coverage}. A geometric abstraction layer then emerged with envelope curve and calculation plane constructions: Zuo \emph{et al.}~\cite{zuo2020envelope_coverage} cast coverage as an envelope curve problem on a sensor-fixed plane, whereas Han \emph{et al.}~\cite{han2021novel_visibility} generalized visibility computation to arbitrary sensor fields through a calculation plane point-in-region test. The present paper specializes the latter construction to side-looking SAR. Recent regional target studies refined this reduction through minimum observation sampling~\cite{jiang2022minimum_observation}, arbitrary attitude maneuver paths~\cite{wang2022echo}, semi-analytical area target visibility~\cite{zhang2022semi_analytical_visibility}, elevation view elements~\cite{gu2024elevation_view}, triple time interval hybridization~\cite{shan2024ttihs}, and the SAR-oriented visible time window method of Shi \emph{et al.}~\cite{shi2025vtwsar}. At the constellation level, multi-satellite scheduling frameworks for very large area observation~\cite{li2025multi_sat_large_area,wang2025aeos_scheduling} and analytic SAR revisit time determination and repeating ground track orbit optimization~\cite{phung2023sar_rgt_orbit} have further advanced the operational use of geometry-driven planning. However, these scheduling models treat visibility as a binary predicate and do not test whether the resulting SAR image quality meets application-level requirements. Chatterjee and Tharmarasa~\cite{chatterjee2024multistage} introduced a multi-stage optimization framework for satellite scheduling over large areas of interest, yet the coupling between observation geometry, echo simulation, and per-window image quality remains absent from all existing frameworks. The main unresolved point is explicit enforcement of side-looking SAR admissibility, rather than generic geometric accessibility, coupled with signal-level quality qualification throughout the window construction process.

Per-window image quality evaluation has received far less attention than visibility computation. Realistic SAR raw data generation rests on accurate range history modeling, which has evolved from classical hyperbolic and equivalent squint formulations through higher-order polynomial parameterizations to the numerically evaluated range models required for high-resolution, long-aperture geometries~\cite{curlander1991synthetic,cumming2005digital,li2023raw_data_sim,li2024accurate_range}, and the choice of focusing algorithm directly governs the fidelity of the resulting image. Classical frequency-domain algorithms, including Range-Doppler (RDA), Chirp Scaling (CSA), and Omega-k, approximate the range history to achieve high throughput~\cite{curlander1991synthetic,cumming2005digital}, whereas time-domain back projection (BP) forms each pixel through exact per-pixel two-way delay computation~\cite{curlander1991synthetic} and recent surveys confirm that this geometric fidelity justifies its higher per-pixel cost in quality-governed decisions~\cite{cruz2022sar_review}. Canonical point target figures of merit, namely impulse response width (IRW), peak sidelobe ratio (PSLR), and integrated sidelobe ratio (ISLR), are well established~\cite{curlander1991synthetic,cumming2005digital}, yet most preprocessing pipelines treat all geometry-feasible windows as equivalent without testing BP-focused image quality against these metrics. The connection between BP-based signal-level quality screening and systematic window evaluation remains unexploited.

To address these gaps, this paper develops an efficient framework for regional SAR visible window generation and per-window quality assessment. The framework constructs geometry-feasible observation windows through three stages: angular bandpass screening, a planar characteristic curve containment test, and one-dimensional boundary bisection. Side-looking SAR admissibility is enforced throughout the preprocessing chain. Each geometry-feasible window is then evaluated by a companion point target stripmap simulation, and only windows satisfying prescribed range and azimuth IRW, PSLR, and ISLR criteria are retained.

The main contributions of this paper are as follows:
\begin{itemize}
  \item A three-stage fast effective window construction procedure for regional targets under side-looking SAR admissibility. The procedure combines coarse angular bandpass screening, a planar characteristic curve containment test on the sensor calculation plane, and one-dimensional boundary bisection that resolves entry and exit epochs to subsecond precision.
  \item A per-window signal-level quality screening procedure using a companion point target evaluated through the same echo generation and back projection (BP) imaging chain as the area scene. Range and azimuth IRW, PSLR, and ISLR are measured against mission-dependent thresholds to retain only windows with acceptable focusing quality.
\end{itemize}

The remainder of this paper is organized as follows. Section~\ref{sec:window_generation} develops the efficient visible window generation method, including coarse angular screening, characteristic curve containment, and boundary bisection refinement. Section~\ref{sec:geosot_evaluation} presents the per-window signal level quality evaluation and effective window screening procedure. Section~\ref{sec:experiments} reports two experiments: an STK-referenced window generation validation and a three-window signal-level quality study. Section~\ref{sec:conclusion} concludes the paper.

\section{Efficient Visible Window Generation}
\label{sec:window_generation}

This section develops the method for constructing geometry feasible candidate observation windows from continuous time satellite visibility. The procedure combines coarse angular screening in the Earth centered Earth fixed frame, refined projection plane intersection testing on the sensor calculation plane, and boundary bisection for precise entry and exit time determination. Under a SAR field of view description aligned with the Systems Tool Kit (STK) side looking payload model~\cite{ansys_stk_ephemeris_file}, continuous time regional visibility is sampled and screened into a sparse set of candidate observation windows.

\subsection{Coarse to Fine Geometric Screening}
\label{sec:coarse_screening}

\begin{figure}
\centering
\includegraphics[width=0.40\textwidth]{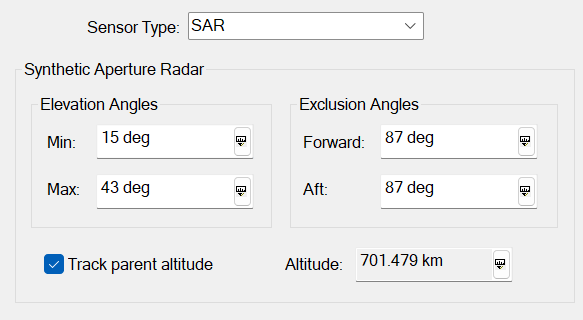}
\caption{STK side looking SAR parameter panel used to define the elevation and exclusion constraints.}
\label{fig:stk_sar_ui}
\end{figure}

Let the regional target boundary polygon be denoted by $\mathcal{G}$. Its boundary vertices and regional centroid are represented by Earth centered Earth fixed (ECEF) position vectors $\{\bm{r}_{g,i}\}_{i=1}^{N}$ and $\bm{r}_{g,c}$, respectively. Candidate window generation comprises a coarse orbit filtering stage and a refined geometric stage under the STK side looking SAR template illustrated in Fig.~\ref{fig:stk_sar_ui}. The minimum and maximum elevation entries bound $\theta_{\mathrm{el},\min}$ and $\theta_{\mathrm{el},\max}$, whereas the forward and aft exclusion entries supply the forward and backward Doppler cone half angles $\theta_{\mathrm{DC,fwd}}$ and $\theta_{\mathrm{DC,bwd}}$, with the aft referenced backward limit restored through the supplementary angle. Because STK evaluates the cone with the satellite velocity expressed in the Earth centered inertial (ECI) frame and then projected into the sensor frame, the satellite velocity $\bm{v}_s$ resolved in the ECEF frame after TEME rotation is distinguished here from the satellite velocity $\bm{v}_{s,\mathrm{I}}$ expressed in the ECI frame for Doppler cone evaluation. The coarse stage retains orbit arcs satisfying angular bandpass constraints on the cross track geocentric angle $\theta_{\mathrm{SOP}}$ and on $\theta_{\mathrm{DC}}$.

Throughout this subsection the WGS-84 reference ellipsoid is adopted as the Earth model.  Satellite and ground positions are expressed as geocentric vectors in the ECEF frame: $\bm{r}_s(t)$ denotes the satellite position at epoch~$t$, $\bm{v}_s(t)$ denotes the satellite velocity in the ECEF frame obtained after frame rotation, and $\bm{r}_g$ denotes the ECEF position of a ground point or polygon vertex.  For geodetic latitude~$\varphi$, longitude~$\lambda$, and ellipsoidal height~$h$, the standard WGS-84 conversion is
\begin{equation}
\bm{r}_g =
\begin{bmatrix}
\left(N_{\!e}+h\right)\cos\varphi\,\cos\lambda\\
\left(N_{\!e}+h\right)\cos\varphi\,\sin\lambda\\
\left[N_{\!e}\!\left(1-e^2\right)+h\right]\sin\varphi
\end{bmatrix},
\quad
N_{\!e}=\frac{a}{\sqrt{1-e^2\sin^2\!\varphi}},
\label{eq:geodetic_to_ecef}
\end{equation}
where $N_{\!e}$ is the prime vertical radius of curvature, $a=6\,378\,137$\,m is the semi-major axis, and $e^2=6.694\,379\,990\,14\times10^{-3}$ is the first eccentricity squared of the WGS-84 ellipsoid.  Orbital states are obtained by feeding North American Aerospace Defense Command (NORAD) Two-Line Element (TLE) sets into the Simplified General Perturbations~4 (SGP4) propagator with the WGS-84 gravity constants, which outputs the position $\bm{r}_{\mathrm{TEME}}$ and velocity $\bm{v}_{\mathrm{TEME}}$ in the True Equator Mean Equinox (TEME) frame.  The TEME vectors are then rotated to the pseudo Earth fixed (PEF) frame by a single axis rotation through the Greenwich Mean Sidereal Time (GMST) angle $\theta_{\mathrm{GMST}}$:
\begin{equation}
\begin{gathered}
\bm{r}_s = \bm{R}_3(\theta_{\mathrm{GMST}})\,\bm{r}_{\mathrm{TEME}},\\
\bm{v}_s = \bm{R}_3(\theta_{\mathrm{GMST}})\,\bm{v}_{\mathrm{TEME}}
         - \bm{\omega}_{\oplus}\times\bm{r}_s,
\label{eq:teme_to_ecef}
\end{gathered}
\end{equation}
where $\bm{R}_3(\cdot)$ denotes a rotation about the $z$-axis, $\bm{\omega}_{\oplus}=(0,0,\omega_{\oplus})^{\!\top}$ with $\omega_{\oplus}=7.292\,115\times10^{-5}$\,rad\,s$^{-1}$, and $\theta_{\mathrm{GMST}}$ is evaluated from the IAU expression given in \cite{vallado2013fundamentals}.  Accordingly, the SGP4-derived TEME states are paired here with the GMST rotation alone \cite{vallado2006revisiting}.

\begin{figure}
\centering
\begin{subfigure}[t]{0.46\textwidth}
\centering
\includegraphics[width=\textwidth]{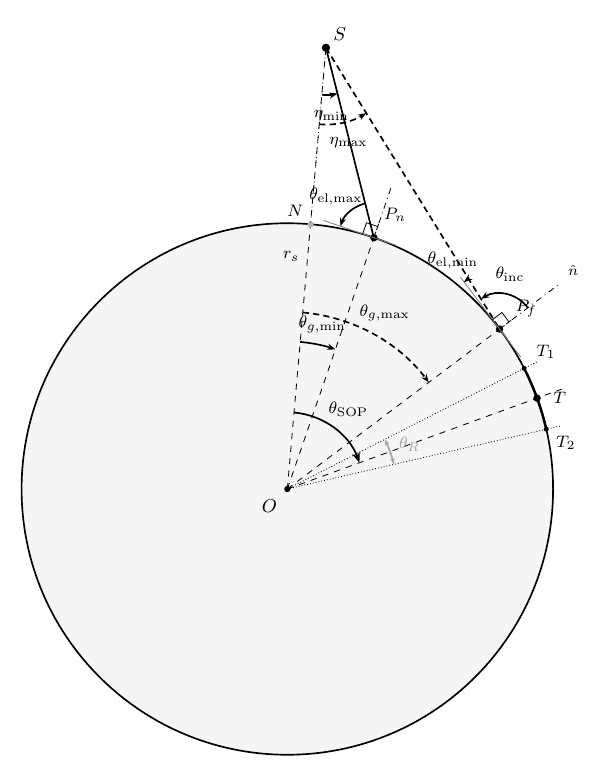}
\caption{Cross track geometry defined by $\theta_{\mathrm{SOP}}$.}
\label{fig:jsvf_theta_sop_2d}
\end{subfigure}
\hfill
\begin{subfigure}[t]{0.46\textwidth}
\centering
\includegraphics[width=\textwidth]{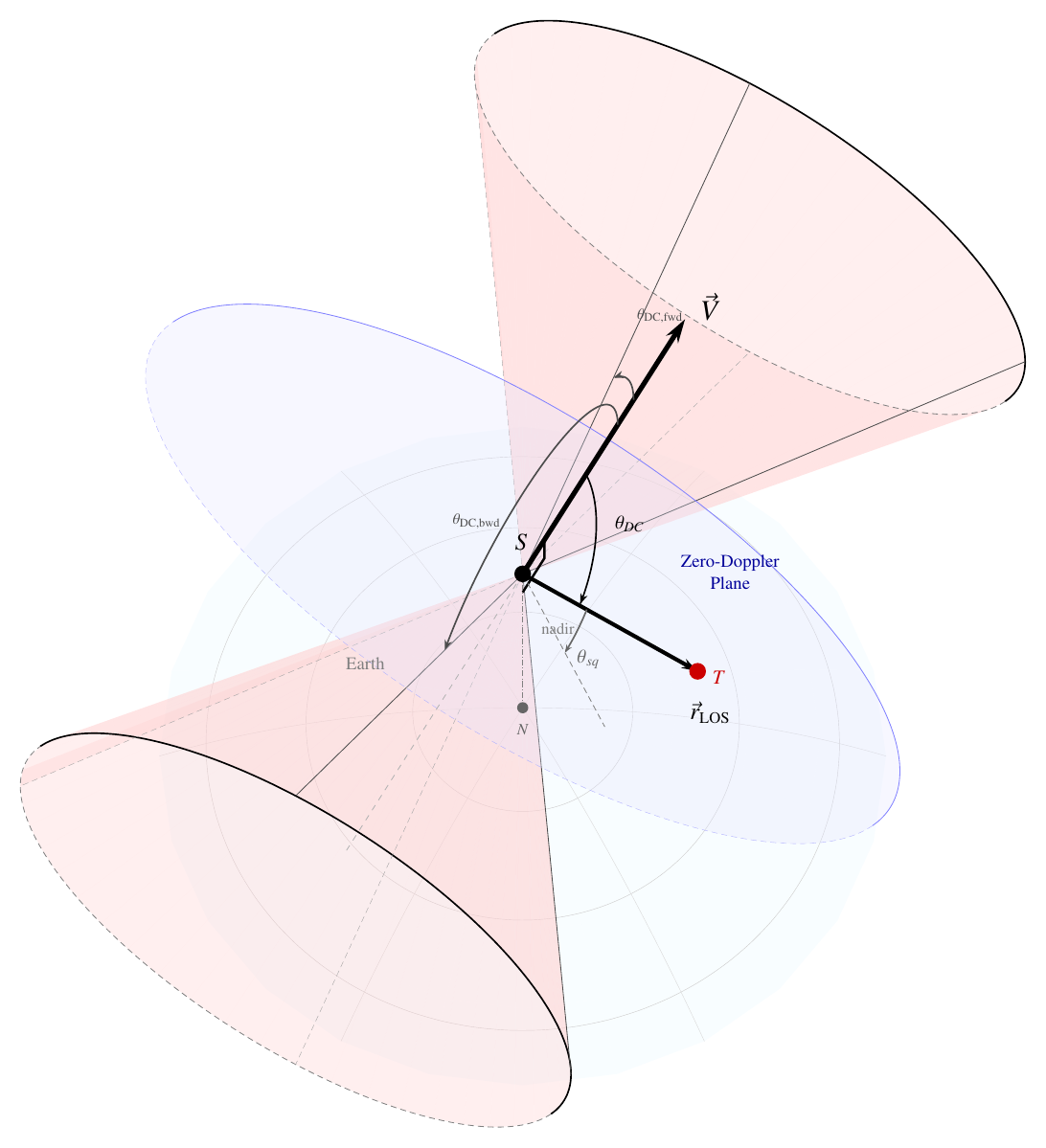}
\caption{Along track geometry defined by $\theta_{\mathrm{DC}}$.}
\label{fig:doppler_cone_angle_3d}
\end{subfigure}
\caption{Geometric components of the coarse angular screening.}
\label{fig:jsvf_geometry}
\end{figure}

Fig.~\ref{fig:jsvf_geometry}\subref{fig:jsvf_theta_sop_2d} illustrates the cross track geometry on a two dimensional Earth centered section. The satellite position is represented by the vector $\bm{r}_s$, and its geocentric orbital radius is denoted by $r_s$. Under the STK side looking payload parameterization, the admissible ground swath is bounded by the near range point~$P_n$ and the far range point~$P_f$. The elevation angle $\theta_{\mathrm{el}}$ is defined at the ground point as the angle between the local horizon and the satellite line of sight. The corresponding radar incidence angle satisfies
\begin{equation}
  \theta_{\mathrm{inc}} = 90^{\circ} - \theta_{\mathrm{el}}.
  \label{eq:el_inc_relation}
\end{equation}
A larger elevation angle corresponds to a smaller geocentric angle.  Accordingly, the near range boundary is governed by the upper elevation limit, whereas the far range boundary is governed by the lower elevation limit.  Considering the Earth centered triangle formed by the Earth center, the satellite, and a ground boundary point, the interior angle at the ground vertex is given by the sum of a right angle and the ground point elevation angle.  Applying the sine rule then yields the following geocentric angular bounds:
\begin{equation}
\begin{aligned}
\theta_{g,\min}&=\bigl(90^{\circ}-\theta_{\mathrm{el},\max}\bigr)
  -\arcsin\!\Bigl(\frac{R_e}{r_s}\cos\theta_{\mathrm{el},\max}\Bigr),\\[4pt]
\theta_{g,\max}&=\bigl(90^{\circ}-\theta_{\mathrm{el},\min}\bigr)
  -\arcsin\!\Bigl(\frac{R_e}{r_s}\cos\theta_{\mathrm{el},\min}\Bigr),
\end{aligned}
\label{eq:elevation_ground_limits}
\end{equation}
where $R_e=6\,371$\,km is the mean Earth radius and the $\arcsin$ argument follows from the sine rule identity
\begin{equation}
  \sin\eta = \frac{R_e}{r_s}\cos\theta_{\mathrm{el}},
  \label{eq:off_nadir_sine_rule}
\end{equation}
with $\eta$ denoting the off nadir angle at the satellite.

The centroid geodetic coordinates are obtained as the arithmetic mean of the vertex coordinates,
\begin{equation}
\varphi_c = \frac{1}{N}\sum_{i=1}^{N}\varphi_i,\qquad
\lambda_c = \frac{1}{N}\sum_{i=1}^{N}\lambda_i,
\label{eq:centroid_geodetic}
\end{equation}
and the centroid position vector $\bm{r}_{g,c}$ is obtained by converting $(\varphi_c,\lambda_c)$ to ECEF coordinates through Eq.~\eqref{eq:geodetic_to_ecef}.  The geocentric half-angle enclosing the polygon and the satellite centroid geocentric angle are
\begin{equation}
\begin{gathered}
\theta_{R}
=
\max_{1\le i\le N}\arccos\!\left(
\frac{\bm{r}_{g,i}\cdot\bm{r}_{g,c}}
{\left\|\bm{r}_{g,i}\right\|\left\|\bm{r}_{g,c}\right\|}
\right),\\
\theta_{\mathrm{SOP}}
=
\arccos\!\left(
\frac{\bm{r}_s\cdot\bm{r}_{g,c}}
{\left\|\bm{r}_s\right\|\left\|\bm{r}_{g,c}\right\|}
\right).
\label{eq:theta_sop}
\end{gathered}
\end{equation}
where $\bm{r}_{g,i}$ is the ECEF position vector of boundary vertex $i$, $\bm{r}_{g,c}$ is the ECEF position vector of the regional centroid, and $\bm{r}_s$ is the satellite position vector at the evaluation epoch.
The cross track bandpass retains arcs satisfying
\begin{equation}
\theta_{g,\min}-\theta_R-\epsilon_m
\;\le\;
\theta_{\mathrm{SOP}}
\;\le\;
\theta_{g,\max}+\theta_R+\epsilon_m,
\label{eq:jsvf_bandpass}
\end{equation}
Inequality~\eqref{eq:jsvf_bandpass} includes an additive margin for altitude variation and discretization.
The Doppler cone angle and its squint companion angle are assembled as follows:
\begin{equation}
\begin{gathered}
\bm{v}_{s,\mathrm{I}}=\bm{v}_s+\bm{\omega}_{\oplus}\times\bm{r}_s,\\
\theta_{\mathrm{DC}}
=
\arccos\!\left(
\frac{\bm{v}_{s,\mathrm{I}}\cdot(\bm{r}_{g,c}-\bm{r}_s)}
{\left\|\bm{v}_{s,\mathrm{I}}\right\|\left\|\bm{r}_{g,c}-\bm{r}_s\right\|}
\right),\\
\theta_{\mathrm{sq}}=90^{\circ}-\theta_{\mathrm{DC}}.
\label{eq:doppler_cone}
\end{gathered}
\end{equation}
where $\bm{v}_s$ is the satellite velocity in the ECEF frame supplied by Eq.~\eqref{eq:teme_to_ecef}, $\bm{v}_{s,\mathrm{I}}$ is the corresponding satellite velocity in the ECI frame, $\theta_{\mathrm{DC}}$ is the Doppler cone angle measured with respect to the centroid line of sight, and $\theta_{\mathrm{sq}}$ is the associated squint angle recorded for later use.
Accordingly, $\bm{v}_s$ is used for kinematics in the rotating Earth model, whereas $\bm{v}_{s,\mathrm{I}}$ is used for Doppler evaluation because the cone definition follows the satellite velocity in the ECI frame rather than the satellite velocity in the ECEF frame.
The along track bandpass is
\begin{equation}
\theta_{\mathrm{DC,fwd}}-\epsilon_m
\;\le\;
\theta_{\mathrm{DC}}
\;\le\;
\theta_{\mathrm{DC,bwd}}+\epsilon_m,
\label{eq:doppler_cone_bandpass}
\end{equation}
where $\theta_{\mathrm{DC,fwd}}$ and $\theta_{\mathrm{DC,bwd}}$ are the forward and backward Doppler cone limits shown in Fig.~\ref{fig:jsvf_geometry}\subref{fig:doppler_cone_angle_3d}.  Only arcs satisfying both Eqs.~\eqref{eq:jsvf_bandpass} and~\eqref{eq:doppler_cone_bandpass} are retained after the coarse angular screening.

\subsection{Characteristics of the SAR Field of View on the Sensor Calculation Plane}

After coarse angular screening, the retained intervals correspond to shortened orbit arcs, so refined screening is carried out on the sensor calculation plane. With Doppler cone limits imposed, the field of view envelope is determined by hyperbolic boundaries in the flight direction and by inner and outer radial bounds set by the minimum and maximum slant ranges in the orthogonal direction. Straight segment approximation would introduce nonnegligible geometric errors at geostationary orbital altitudes, where the illuminated area can extend over several hundred kilometres.

\begin{figure*}
\centering
\begin{subfigure}[t]{0.8\textwidth}
\centering
\includegraphics[width=\linewidth]{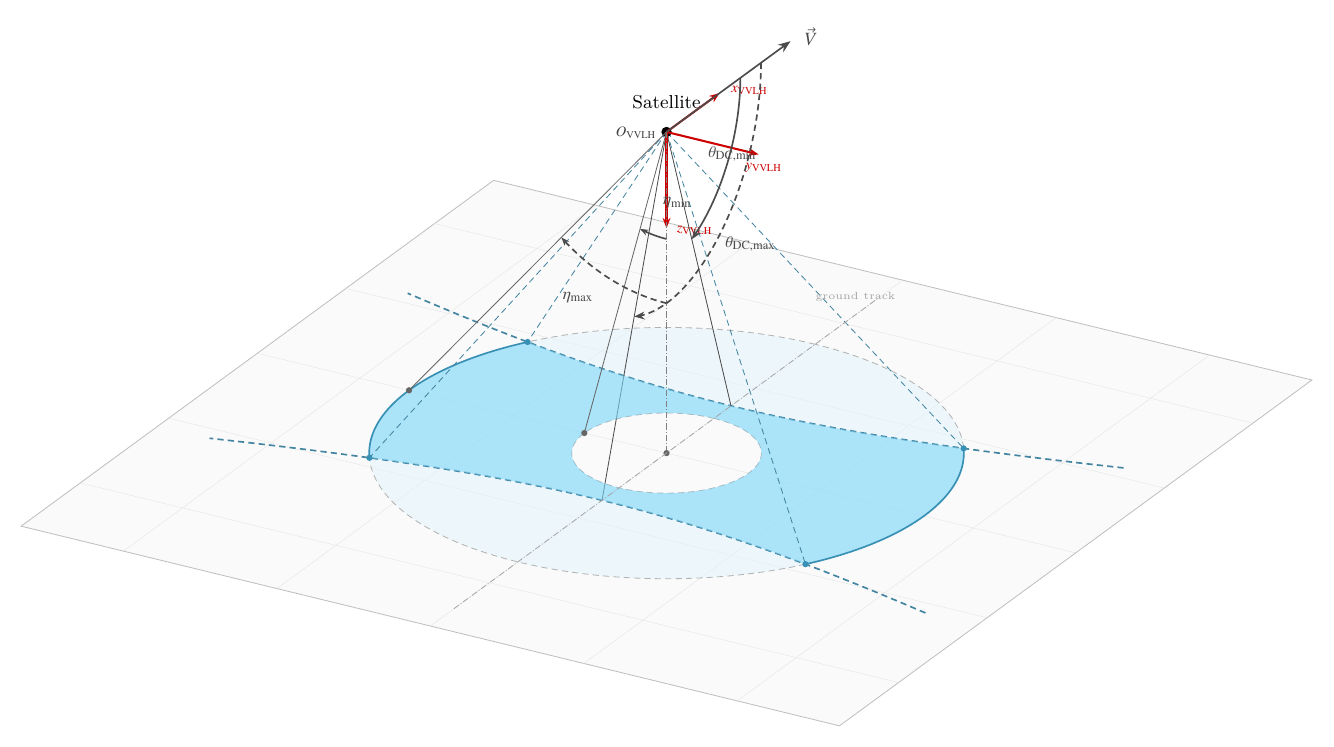}
\caption{Annular field of view envelope on the ground plane.}
\label{fig:stk_fov_envelope}
\end{subfigure}
\hfill
\begin{subfigure}[t]{0.6\textwidth}
\centering
\includegraphics[width=\linewidth]{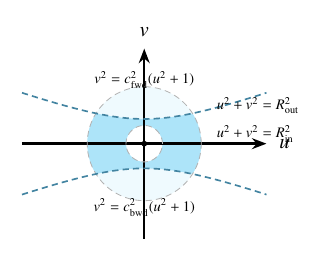}
\caption{Characteristic boundary in the $(u,v)$ plane.}
\label{fig:calc_plane_fov}
\end{subfigure}
\caption{SAR field of view as an annular envelope on the ground plane \subref{fig:stk_fov_envelope} and as an analytic boundary on the VVLH calculation plane \subref{fig:calc_plane_fov}.}
\label{fig:fov_dual_view}
\end{figure*}

A closed form description of the curved boundary follows from the calculation plane approach developed for arbitrary sensor footprints in Ref.~\cite{han2021novel_visibility}, specialized here to the side looking SAR envelope after Doppler cone clipping. Each candidate ground point is mapped into a plane fixed to the sensor, so visibility reduces to membership in a region bounded by a characteristic curve that is independent of time. The calculation plane is defined as the plane $z=1$ in the vehicle vertical local horizontal (VVLH) frame. As illustrated in Fig.~\ref{fig:fov_dual_view}\subref{fig:stk_fov_envelope}, the VVLH triad is attached to the satellite origin $O_{\mathrm{VVLH}}$. The axis $x_{\mathrm{VVLH}}$ carries the along track direction in the imaging convention, $z_{\mathrm{VVLH}}$ points toward nadir, and $y_{\mathrm{VVLH}}$ follows from the right handed construction in Eq.~\eqref{eq:vvlh_triad}. Because the instantaneous footprint must be tied to orbital kinematics defined in the ECI frame rather than to the satellite velocity $\bm{v}_s$ in the ECEF frame, the normal $\hat{\bm{y}}$ is formed with $\bm{v}_{s,\mathrm{I}}$ from Eq.~\eqref{eq:doppler_cone}:
\begin{equation}
\hat{\bm{z}} = -\frac{\bm{r}_s}{\|\bm{r}_s\|},\quad
\hat{\bm{y}} = -\frac{\bm{r}_s\times\bm{v}_{s,\mathrm{I}}}{\|\bm{r}_s\times\bm{v}_{s,\mathrm{I}}\|},\quad
\hat{\bm{x}} = \hat{\bm{y}}\times\hat{\bm{z}}.
\label{eq:vvlh_triad}
\end{equation}
where $\hat{\bm{x}}$, $\hat{\bm{y}}$, and $\hat{\bm{z}}$ are the orthonormal basis vectors of the VVLH frame.
The axis $\hat{\bm{x}}$ lies in the instantaneous orbital plane and is orthogonal to the position vector. For circular orbits, $\hat{\bm{x}}_{\mathrm{VVLH}}$ coincides with $\bm{v}_{s,\mathrm{I}}$; for noncircular orbits, a finite angle exists between them because of the flight path angle. Because the SAR satellites considered here operate mainly in nearly circular sun synchronous low Earth orbits, $\bm{v}_{s,\mathrm{I}}$ is approximated as aligned with $\hat{\bm{x}}_{\mathrm{VVLH}}$ in what follows. Under this approximation, the Doppler cone axis follows $\hat{\bm{x}}_{\mathrm{VVLH}}$, and the forward and backward boundaries appear on the calculation plane as hyperbolic branches given later by Eqs.~\eqref{eq:dc_constraint_uv_fwd}--\eqref{eq:dc_constraint_uv_bwd}. The rotation from the ECEF frame to the VVLH frame and the projection onto the calculation plane are written as
\begin{equation}
\begin{gathered}
\bm{T} = \begin{bmatrix} \hat{\bm{x}} & \hat{\bm{y}} & \hat{\bm{z}} \end{bmatrix}^{\top},\\
\begin{bmatrix} x_{\mathrm{VVLH}} & y_{\mathrm{VVLH}} & z_{\mathrm{VVLH}} \end{bmatrix}^{\top} = \bm{T}(\bm{r}_g-\bm{r}_s),\\
(u,v) = \left(\frac{y_{\mathrm{VVLH}}}{z_{\mathrm{VVLH}}},\;\frac{x_{\mathrm{VVLH}}}{z_{\mathrm{VVLH}}}\right),
\label{eq:calc_plane_projection}
\end{gathered}
\end{equation}
where $u$ is the cross track coordinate and $v$ the along track coordinate under the perspective normalization in Eq.~\eqref{eq:calc_plane_projection}, expressed in the VVLH frame centered at the spacecraft.  Visibility is thus reduced to deciding whether the projected point lies inside the characteristic curve $\Theta$ of the field of view; $\Theta$ is a fixed closed curve in the $(u,v)$ plane and summarizes the sensor geometry without explicit time dependence.

\begin{figure*}
\centering
\begin{subfigure}[t]{0.31\textwidth}
\centering
\includegraphics[width=\linewidth]{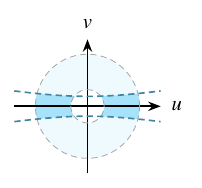}
\caption{$c_{\mathrm{fwd}}\le R_{\mathrm{in}},\;c_{\mathrm{bwd}}\le R_{\mathrm{in}}$.}
\label{fig:fov_case_a}
\end{subfigure}
\hfill
\begin{subfigure}[t]{0.31\textwidth}
\centering
\includegraphics[width=\linewidth]{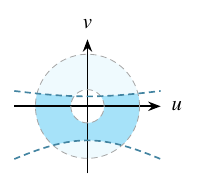}
\caption{$c_{\mathrm{fwd}}\le R_{\mathrm{in}},\;R_{\mathrm{in}}<c_{\mathrm{bwd}}<R_{\mathrm{out}}$.}
\label{fig:fov_case_b}
\end{subfigure}
\hfill
\begin{subfigure}[t]{0.31\textwidth}
\centering
\includegraphics[width=\linewidth]{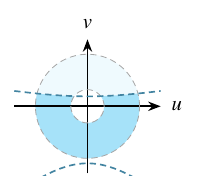}
\caption{$c_{\mathrm{fwd}}\le R_{\mathrm{in}},\;c_{\mathrm{bwd}}\ge R_{\mathrm{out}}$.}
\label{fig:fov_case_c}
\end{subfigure}

\vspace{0.3em}

\begin{subfigure}[t]{0.31\textwidth}
\centering
\includegraphics[width=\linewidth]{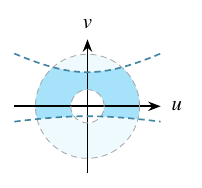}
\caption{$R_{\mathrm{in}}<c_{\mathrm{fwd}}<R_{\mathrm{out}},\;c_{\mathrm{bwd}}\le R_{\mathrm{in}}$.}
\label{fig:fov_case_d}
\end{subfigure}
\hfill
\begin{subfigure}[t]{0.31\textwidth}
\centering
\includegraphics[width=\linewidth]{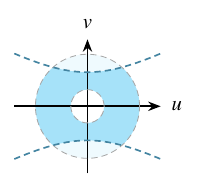}
\caption{$R_{\mathrm{in}}<c_{\mathrm{fwd}}<R_{\mathrm{out}},\;R_{\mathrm{in}}<c_{\mathrm{bwd}}<R_{\mathrm{out}}$.}
\label{fig:fov_case_e}
\end{subfigure}
\hfill
\begin{subfigure}[t]{0.31\textwidth}
\centering
\includegraphics[width=\linewidth]{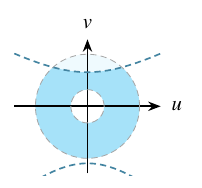}
\caption{$R_{\mathrm{in}}<c_{\mathrm{fwd}}<R_{\mathrm{out}},\;c_{\mathrm{bwd}}\ge R_{\mathrm{out}}$.}
\label{fig:fov_case_f}
\end{subfigure}

\vspace{0.3em}

\begin{subfigure}[t]{0.31\textwidth}
\centering
\includegraphics[width=\linewidth]{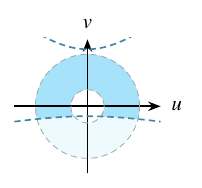}
\caption{$c_{\mathrm{fwd}}\ge R_{\mathrm{out}},\;c_{\mathrm{bwd}}\le R_{\mathrm{in}}$.}
\label{fig:fov_case_g}
\end{subfigure}
\hfill
\begin{subfigure}[t]{0.31\textwidth}
\centering
\includegraphics[width=\linewidth]{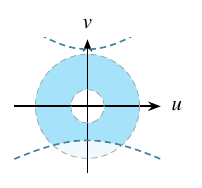}
\caption{$c_{\mathrm{fwd}}\ge R_{\mathrm{out}},\;R_{\mathrm{in}}<c_{\mathrm{bwd}}<R_{\mathrm{out}}$.}
\label{fig:fov_case_h}
\end{subfigure}
\hfill
\begin{subfigure}[t]{0.31\textwidth}
\centering
\includegraphics[width=\linewidth]{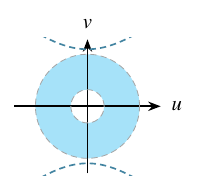}
\caption{$c_{\mathrm{fwd}}\ge R_{\mathrm{out}},\;c_{\mathrm{bwd}}\ge R_{\mathrm{out}}$.}
\label{fig:fov_case_i}
\end{subfigure}
\caption{Nine ordered configurations of the admissible domain $\mathcal{Q}$ on the $(u,v)$ plane, classified by the relative positions of $c_{\mathrm{fwd}}$ and $c_{\mathrm{bwd}}$ with respect to $R_{\mathrm{in}}$ and $R_{\mathrm{out}}$.}
\label{fig:calc_plane_fov_cases}
\end{figure*}

For side looking SAR the characteristic curve admits a concise analytic description; the geometry is illustrated in Fig.~\ref{fig:fov_dual_view}\subref{fig:calc_plane_fov}.  Fig.~\ref{fig:fov_dual_view}\subref{fig:stk_fov_envelope} and Fig.~\ref{fig:jsvf_geometry}\subref{fig:jsvf_theta_sop_2d} jointly show that the radial coordinate on the calculation plane is fixed by the off nadir angle at the satellite.  Limits on elevation at the ground vertex imply corresponding bounds on off nadir angle at the sensor through Eq.~\eqref{eq:off_nadir_sine_rule}.  The larger elevation angle fixes the near range ground boundary and the smaller angle fixes the far range ground boundary; the resulting limits on off nadir angle at the sensor therefore determine the inner and outer radii of the annular admissible domain on the calculation plane, as recorded in Eq.~\eqref{eq:annular_radii_eta}.
\begin{equation}
R_{\mathrm{in}} = \tan\eta_{\min},\qquad
R_{\mathrm{out}} = \tan\eta_{\max}.
\label{eq:annular_radii_eta}
\end{equation}
Substituting Eq.~\eqref{eq:off_nadir_sine_rule} into Eq.~\eqref{eq:annular_radii_eta} then yields the explicit annular bounds
\begin{equation}
\begin{gathered}
R_{\mathrm{in}}^2 \;\le\; u^2 + v^2 \;\le\; R_{\mathrm{out}}^2,\\
R_{\mathrm{in}} = \frac{R_g\cos\theta_{\mathrm{el},\max}}
  {\sqrt{r_s^2-R_g^2\cos^2\!\theta_{\mathrm{el},\max}}},\\
R_{\mathrm{out}} = \frac{R_g\cos\theta_{\mathrm{el},\min}}
  {\sqrt{r_s^2-R_g^2\cos^2\!\theta_{\mathrm{el},\min}}},
\label{eq:annular_constraint_uv}
\end{gathered}
\end{equation}
where $R_g$ denotes the geocentric distance of the ground point.  In the coarse screening of Section~\ref{sec:coarse_screening} the mean spherical radius $R_e$ suffices; for refined screening on the ellipsoid, $R_g$ is obtained from Eq.~\eqref{eq:geodetic_to_ecef} at the target centroid latitude with zero height.  Each Doppler limit admits the usual interpretation as a right circular cone with apex at the satellite and axis aligned with $x_{\mathrm{VVLH}}$ in Fig.~\ref{fig:fov_dual_view}\subref{fig:stk_fov_envelope}.  Because the calculation plane is orthogonal to $z_{\mathrm{VVLH}}$, it is parallel to the cone axis.  Elementary conic section geometry implies that such a plane cuts the cone in a hyperbola.  The forward and backward Doppler limits therefore appear as two branches of one hyperbolic family on the calculation plane, as in Fig.~\ref{fig:fov_dual_view}\subref{fig:calc_plane_fov}, which yields the following single sided bounds:
\begin{equation}
v \le c_{\mathrm{fwd}}\sqrt{u^2+1},\qquad
c_{\mathrm{fwd}} = |\cot\theta_{\mathrm{DC,min}}|,
\label{eq:dc_constraint_uv_fwd}
\end{equation}
\begin{equation}
v \ge -c_{\mathrm{bwd}}\sqrt{u^2+1},\qquad
c_{\mathrm{bwd}} = |\cot\theta_{\mathrm{DC,max}}|.
\label{eq:dc_constraint_uv_bwd}
\end{equation}
where $c_{\mathrm{fwd}}$ and $c_{\mathrm{bwd}}$ denote the forward and backward Doppler cone coefficients, respectively, and $\theta_{\mathrm{DC,min}}$ and $\theta_{\mathrm{DC,max}}$ are the corresponding forward and backward bounds on the Doppler cone angle.  The admissible domain $\mathcal{Q}$ is the intersection of the inner and outer radial annuli with these two hyperbolic half plane constraints.  Because $c_{\mathrm{fwd}}$ and $c_{\mathrm{bwd}}$ are ordered independently relative to $R_{\mathrm{in}}$ and $R_{\mathrm{out}}$, nine distinct relative configurations arise, as shown in Fig.~\ref{fig:calc_plane_fov_cases}.  They differ only in topological connectivity and in which segments of the boundary remain active; the case structure is precisely what is required by the refined intersection test between the regional polygon and $\mathcal{Q}$ developed next.

\subsection{Refined Screening Based on the Characteristic Curve FOV}
\label{sec:ccpp_refined}

\begin{figure}
\centering
\begin{subfigure}[t]{0.31\textwidth}
\centering
\includegraphics[width=\textwidth]{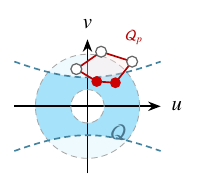}
\caption{VertexContainment.}
\label{fig:test_vc}
\end{subfigure}
\hfill
\begin{subfigure}[t]{0.31\textwidth}
\centering
\includegraphics[width=\textwidth]{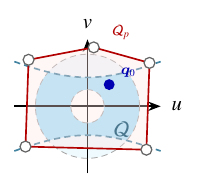}
\caption{ReverseContainment.}
\label{fig:test_rc}
\end{subfigure}
\hfill
\begin{subfigure}[t]{0.31\textwidth}
\centering
\includegraphics[width=\textwidth]{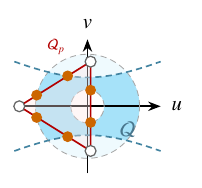}
\caption{BoundaryCrossing.}
\label{fig:test_bc}
\end{subfigure}
\caption{Three complementary geometric tests for the polygon domain intersection predicate~\eqref{eq:ccpp_predicate}. }
\label{fig:three_tests}
\end{figure}

The preceding derivation yields a fixed analytic admissible domain $\mathcal{Q}$ on the $(u,v)$ plane and identifies its possible topological classes. The regional visibility test becomes a polygon domain intersection problem, and the refined screening strategy is defined directly by the active boundary components of $\mathcal{Q}$. The refined screening predicate requires nonempty intersection between the projected regional polygon and the admissible domain on the characteristic $(u,v)$ plane:
\begin{equation}
\mathcal{Q}_p(\mathcal{G},t) \cap \mathcal{Q} \neq \varnothing,
\label{eq:ccpp_predicate}
\end{equation}
Projected vertices are constructed by the map in Eq.~\eqref{eq:calc_plane_projection}.  The predicate is evaluated by three complementary geometric tests, illustrated schematically in Fig.~\ref{fig:three_tests}.

\textbf{VertexContainment and ReverseContainment.}
The first test checks each projected vertex against the annular and Doppler inequalities in Eqs.~\eqref{eq:annular_constraint_uv}--\eqref{eq:dc_constraint_uv_bwd}; it is topology independent and detects any partial overlap in which at least one polygon vertex falls inside the admissible domain.  However, when the regional target is large enough for its projection to surround that domain entirely, as illustrated in Fig.~\ref{fig:test_rc}, every vertex lies outside the domain and the first test yields a false negative.  The reverse containment test eliminates this gap by inverting the containment query.  For the connected configurations shown in Fig.~\ref{fig:calc_plane_fov_cases}\subref{fig:fov_case_b}--\subref{fig:fov_case_i}, a deterministic representative point is fixed on the inner radius along the horizontal axis on the positive side; it lies inside the admissible domain by construction.  For the disjoint lobe configuration~a, one representative per lobe is used, placed on the inner radius at the negative and positive horizontal positions, respectively.  The projected polygon vertices are traversed in cyclic order.  From each representative, shoot a horizontal half ray toward increasing abscissa and apply the odd even rule for point-in-polygon classification~\cite{orourke1994comp_geom}; the compact indicator form is Eq.~\eqref{eq:rc_odd_even}.
\begin{equation}
I_{\mathrm{RC}}(\bm{q}_0,\mathcal{Q}_p)=
\begin{cases}
1, & N_{\mathrm{cross}}(\bm{q}_0,\mathcal{Q}_p) \bmod 2 = 1,\\
0, & N_{\mathrm{cross}}(\bm{q}_0,\mathcal{Q}_p) \bmod 2 = 0,
\end{cases}
\label{eq:rc_odd_even}
\end{equation}
where $\bm{q}_0$ is the representative point and $N_{\mathrm{cross}}(\bm{q}_0,\mathcal{Q}_p)$ is the number of intersections between the half ray emitted from $\bm{q}_0$ and the boundary of the projected polygon.
Unity of the indicator corresponds to the representative lying inside the projected polygon; the crossing count tallies intersections between that half ray and polygon edges under the usual half open convention so that vertex touching cases are not double counted.  With these fixed representative choices, ReverseContainment becomes completely deterministic and requires no additional representative point search; a step-by-step implementation appears in Appendix~\ref{app:reverse_containment} and Algorithm~\ref{alg:reverse_containment_appendix}.

\textbf{BoundaryCrossing.}
The third test detects the remaining case in which the projected polygon and $\mathcal{Q}$ overlap without either fully containing the other, as illustrated in Fig.~\ref{fig:test_bc}.  Each polygon edge is tested against the active boundary arcs of $\mathcal{Q}$.  The boundary $\partial\mathcal{Q}$ comprises at most four curve families: the inner and outer circular arcs from Eq.~\eqref{eq:annular_constraint_uv}, and the forward and backward hyperbolic arcs from Eqs.~\eqref{eq:dc_constraint_uv_fwd} and~\eqref{eq:dc_constraint_uv_bwd}.  The activation condition of each hyperbolic branch follows directly from Eqs.~\eqref{eq:dc_constraint_uv_fwd} and~\eqref{eq:dc_constraint_uv_bwd}, while the hyperbola circle endpoint abscissa is given by Eq.~\eqref{eq:hyp_circle_endpoint}:
\begin{equation}
u_{\star} = \pm\,\sqrt{\frac{R^2 - c^2}{1 + c^2}},
\qquad R \ge c.
\label{eq:hyp_circle_endpoint}
\end{equation}
For each polygon edge, the intersection with each active boundary curve of $\partial\mathcal{Q}$ is formulated as a quadratic subproblem.  The edge parametrization is
\begin{equation}
\begin{gathered}
u(\lambda)=u_{s}+\lambda\,\Delta u,\qquad
v(\lambda)=v_{s}+\lambda\,\Delta v,\\
\Delta u=u_{e}-u_{s},\quad \Delta v=v_{e}-v_{s},\quad 0\le \lambda\le 1.
\label{eq:edge_parametric_line}
\end{gathered}
\end{equation}
where $(u_s,v_s)$ and $(u_e,v_e)$ are the start and end points of the polygon edge, and $\lambda$ is the edge parameter.
Substitution of Eq.~\eqref{eq:edge_parametric_line} into each active boundary relation yields a quadratic of the unified form
\begin{equation}
A\,\lambda^{2}+B\,\lambda+C=0,
\label{eq:edge_quadratic_generic}
\end{equation}
where the coefficients $A$, $B$, and $C$ for the four boundary curve families are summarized in Table~\ref{tab:bc_method}.  Because the hyperbolic relations are squared during elimination, every candidate root must additionally satisfy the validation conditions listed in Table~\ref{tab:bc_method}.  For each quadratic, the candidate roots are
\begin{equation}
\lambda^{*}=\frac{-B\pm\sqrt{B^{2}-4AC}}{2A}.
\label{eq:quadratic_root}
\end{equation}
If the radicand is negative, no real root exists and the corresponding curve contributes no intersection for that edge.  Each real root $\lambda^*$ yields a recovered point $(u^*,v^*)$ via Eq.~\eqref{eq:edge_parametric_line}.  A root is accepted as a genuine edge boundary intersection only if all validation conditions listed in Table~\ref{tab:bc_method} are satisfied.  A valid intersection on any active curve suffices for the BoundaryCrossing predicate to return true.

\begin{table*}
\centering
\caption{Quadratic coefficients, activation, and validation rules for the boundary curves of $\partial\mathcal{Q}$.}
\label{tab:bc_method}
\scriptsize
\setlength{\tabcolsep}{3pt}
\renewcommand{\arraystretch}{1.6}
\begin{tabular}{@{}p{0.11\linewidth}p{0.11\linewidth}lllp{0.24\linewidth}@{}}
\toprule
Boundary curve & Activation & $A$ & $B$ & $C$ & Intersection validation \\
\midrule
Inner circle
 & Always active
 & $\Delta u^{2}\!+\!\Delta v^{2}$
 & $2(u_{s}\Delta u\!+\!v_{s}\Delta v)$
 & $u_{s}^{2}\!+\!v_{s}^{2}\!-\!R_{\mathrm{in}}^{2}$
 & $0\!\le\!\lambda^*\!\le\!1$\newline $v^*\!\le\! c_{\mathrm{fwd}}\sqrt{u^{*2}\!+\!1}$\newline $v^*\!\ge\! -c_{\mathrm{bwd}}\sqrt{u^{*2}\!+\!1}$ \\
\midrule
Outer circle
 & Always active
 & $\Delta u^{2}\!+\!\Delta v^{2}$
 & $2(u_{s}\Delta u\!+\!v_{s}\Delta v)$
 & $u_{s}^{2}\!+\!v_{s}^{2}\!-\!R_{\mathrm{out}}^{2}$
 & $0\!\le\!\lambda^*\!\le\!1$\newline $v^*\!\le\! c_{\mathrm{fwd}}\sqrt{u^{*2}\!+\!1}$\newline $v^*\!\ge\! -c_{\mathrm{bwd}}\sqrt{u^{*2}\!+\!1}$ \\
\midrule
Fwd hyperbola
 & $c_{\mathrm{fwd}}\!<\!R_{\mathrm{out}}$
 & $\Delta v^{2}\!-\!c_{\mathrm{fwd}}^{2}\Delta u^{2}$
 & $2(v_{s}\Delta v\!-\!c_{\mathrm{fwd}}^{2}u_{s}\Delta u)$
 & $v_{s}^{2}\!-\!c_{\mathrm{fwd}}^{2}(u_{s}^{2}\!+\!1)$
 & $0\!\le\!\lambda^*\!\le\!1$\newline $R_{\mathrm{in}}^2\!\le\! u^{*2}\!+\!v^{*2}\!\le\! R_{\mathrm{out}}^2$\newline $v^*\!>\!0$ \\
\midrule
Bwd hyperbola
 & $c_{\mathrm{bwd}}\!<\!R_{\mathrm{out}}$
 & $\Delta v^{2}\!-\!c_{\mathrm{bwd}}^{2}\Delta u^{2}$
 & $2(v_{s}\Delta v\!-\!c_{\mathrm{bwd}}^{2}u_{s}\Delta u)$
 & $v_{s}^{2}\!-\!c_{\mathrm{bwd}}^{2}(u_{s}^{2}\!+\!1)$
 & $0\!\le\!\lambda^*\!\le\!1$\newline $R_{\mathrm{in}}^2\!\le\! u^{*2}\!+\!v^{*2}\!\le\! R_{\mathrm{out}}^2$\newline $v^*\!<\!0$ \\
\bottomrule
\end{tabular}
\end{table*}

\begin{figure*}
\centering
\includegraphics[width=0.7\textwidth]{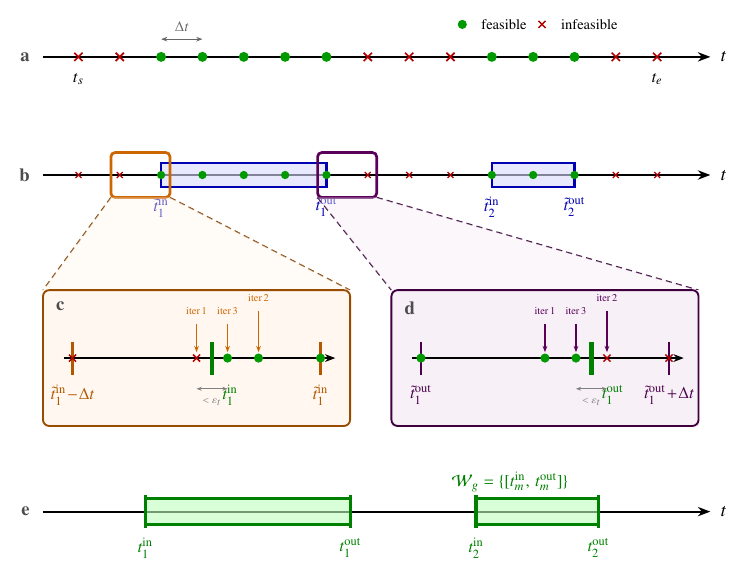}
\caption{Schematic of bisection refinement for coarse boundary brackets.}
\label{fig:bisection_refinement}
\end{figure*}

In the disjoint lobe configuration of Fig.~\ref{fig:fov_case_a}, each validated root must additionally be assigned to the correct connected component.  The number of quadratic solves per edge ranges from two to four depending on which hyperbolic branches are active, as determined by the activation conditions in Table~\ref{tab:bc_method}.

\textbf{Bisection refinement.}
The three tests above provide a boolean predicate at each discrete sampling epoch.  The coarse angular screening stage groups the entire planning horizon into a set of coarse candidate intervals $\mathcal{W}_c=\{[t_{s,i},\,t_{e,i}]\}_{i=1}^{N_c}$; within each such interval, uniform evaluation with the refined step $\Delta t_f$ identifies the feasible epochs.  Grouping contiguous feasible epochs within each element of $\mathcal{W}_c$ yields coarse brackets $[\tilde t^{\mathrm{in}}_m,\,\tilde t^{\mathrm{out}}_m]$ whose boundaries are resolved only to $\Delta t_f$.  To locate the true entry and exit times $t^{\mathrm{in}}_m$ and $t^{\mathrm{out}}_m$ with subsecond precision, a standard one dimensional bisection is applied to predicate~\eqref{eq:ccpp_predicate} at each bracket boundary.  The method starts from a bracket whose two endpoints lie on opposite sides of the feasibility transition, computes the midpoint, re-evaluates Eq.~\eqref{eq:ccpp_predicate} at that midpoint, and then replaces the endpoint with the same feasibility state as the midpoint.  Repeating this update shrinks the bracket monotonically while preserving one feasible and one infeasible endpoint, as illustrated in Fig.~\ref{fig:bisection_refinement}.  For the entry time, the search interval is $[\tilde t^{\mathrm{in}}_m-\Delta t_f,\;\tilde t^{\mathrm{in}}_m]$, bracketing one infeasible and one feasible endpoint.  The exit time is treated symmetrically on $[\tilde t^{\mathrm{out}}_m,\;\tilde t^{\mathrm{out}}_m+\Delta t_f]$, where the feasible end is on the left.  Iteration continues until the interval width falls below a prescribed tolerance $\varepsilon_t$, yielding the refined geometry feasible windows $\mathcal{W}_g=\{[t^{\mathrm{in}}_m,\,t^{\mathrm{out}}_m]\}_{m=1}^{N_w}$.

Algorithm~\ref{alg:ccpp_refined_screening} presents the complete candidate window generation procedure for a single satellite and regional target pair. The input consists of a satellite descriptor $\mathcal{S}$ and a scenario descriptor $\mathcal{X}$. The former specifies the TLE information together with the SAR geometric constraints, whereas the latter defines the planning interval, the regional target boundary, and the numerical tolerances used in screening and boundary refinement. The output is the geometry feasible candidate window set $\mathcal{W}_g$ for the satellite target pair.

\begin{breakablealgorithm}
\caption{Fast coarse to fine candidate window generation for one satellite target pair}
\label{alg:ccpp_refined_screening}
\begin{algorithmic}[1]
\Require Satellite descriptor $\mathcal{S}=\bigl(\mathrm{TLE},\Theta_{\mathrm{SAR}}\bigr)$ with $\Theta_{\mathrm{SAR}}=\bigl(\theta_{\mathrm{el},\min},\theta_{\mathrm{el},\max},\theta_{\mathrm{DC,fwd}},\theta_{\mathrm{DC,bwd}}\bigr)$; scenario descriptor $\mathcal{X}=\bigl([t_0^{\mathrm{UTC}},t_1^{\mathrm{UTC}}],G,\varepsilon_m,\Delta t_c,\Delta t_f,\varepsilon_t\bigr)$ with $G=\{(\varphi_i,\lambda_i)\}_{i=1}^{N}$
\Ensure Geometry feasible candidate window set for the satellite target pair, $\mathcal{W}_g=\{[t^{\mathrm{in}}_m,\,t^{\mathrm{out}}_m]\}_{m=1}^{N_w}$
\State Convert $\{(\varphi_i,\lambda_i)\}_{i=1}^{N}$ to ECEF boundary vectors $\{\bm{r}_{g,i}\}_{i=1}^{N}$ via Eq.~\eqref{eq:geodetic_to_ecef} with $h=0$, and form $\mathcal{G}=\{\bm{r}_{g,i}\}_{i=1}^{N}$
\State Compute the centroid coordinates $(\varphi_c,\lambda_c)$ via Eq.~\eqref{eq:centroid_geodetic}, then convert them to the ECEF centroid vector $\bm{r}_{g,c}$ via Eq.~\eqref{eq:geodetic_to_ecef}
\State $\mathcal{W}^{\mathrm{ep}}_c\leftarrow\varnothing$, $\mathcal{W}_g\leftarrow\varnothing$ \Comment{coarse retained epochs and refined windows}
\For{$t=t_0^{\mathrm{UTC}}$ \textbf{to} $t_1^{\mathrm{UTC}}$ \textbf{step} $\Delta t_c$}
  \State Propagate the TLE with SGP4 and rotate the TEME state to ECEF to obtain $\bm{r}_s(t)$ and $\bm{v}_s(t)$
  \State Compute $\theta_{\mathrm{SOP}}$ and $\theta_{\mathrm{DC}}$ via Eqs.~\eqref{eq:theta_sop} and~\eqref{eq:doppler_cone}
  \If{Eqs.~\eqref{eq:jsvf_bandpass} and~\eqref{eq:doppler_cone_bandpass} are both satisfied}
    \State $\mathcal{W}^{\mathrm{ep}}_c\leftarrow\mathcal{W}^{\mathrm{ep}}_c\cup\{t\}$ \Comment{coarse angular screening, Fig.~\ref{fig:jsvf_geometry}}
  \EndIf
\EndFor
\State Group contiguous epochs in $\mathcal{W}^{\mathrm{ep}}_c$ into coarse candidate intervals $\mathcal{W}_c=\{[t_{s,i},\,t_{e,i}]\}_{i=1}^{N_c}$
\State Compute the geocentric angle bounds $\theta_{g,\min}$ and $\theta_{g,\max}$ from Eq.~\eqref{eq:elevation_ground_limits}
\State Convert the elevation angle limits to off nadir bounds $\eta_{\min}$ and $\eta_{\max}$ via Eq.~\eqref{eq:off_nadir_sine_rule}
\State Compute $R_{\mathrm{in}}=\tan\eta_{\min}$ and $R_{\mathrm{out}}=\tan\eta_{\max}$ from Eq.~\eqref{eq:annular_radii_eta}, then obtain the explicit annular bounds from Eq.~\eqref{eq:annular_constraint_uv}
\State Compute the Doppler cone coefficients $c_{\mathrm{fwd}}$ and $c_{\mathrm{bwd}}$ from Eqs.~\eqref{eq:dc_constraint_uv_fwd}--\eqref{eq:dc_constraint_uv_bwd}
\State Assemble the admissible domain $\mathcal{Q}$ and determine its active boundary branches from Fig.~\ref{fig:calc_plane_fov_cases}
\State $i\leftarrow 1$
\While{$i\leq N_c$}
  \State $\mathcal{W}^{\mathrm{ep}}_f\leftarrow\varnothing$ \Comment{feasible epochs after characteristic curve refinement}
  \For{$t=t_{s,i}$ \textbf{to} $t_{e,i}$ \textbf{step} $\Delta t_f$}
    \State Propagate the TLE with SGP4 and rotate the TEME state to ECEF to obtain $\bm{r}_s(t)$ and $\bm{v}_s(t)$
    \State Project the polygon vertices $\{\bm{r}_{g,i}\}_{i=1}^{N}$ onto the $(u,v)$ plane via Eq.~\eqref{eq:calc_plane_projection} to obtain $\mathcal{Q}_p(\mathcal{G},t)$
    \State $b_1\leftarrow$ \Call{VertexContainment}{$\mathcal{Q}_p,\,\mathcal{Q}$} \Comment{Eqs.~\eqref{eq:annular_constraint_uv}--\eqref{eq:dc_constraint_uv_bwd}, Fig.~\ref{fig:test_vc}}
    \State $b_2\leftarrow$ \Call{ReverseContainment}{$\mathcal{Q}_p,\,\mathcal{Q}$} \Comment{Eq.~\eqref{eq:rc_odd_even}, Fig.~\ref{fig:test_rc}}
    \State $b_3\leftarrow$ \Call{BoundaryCrossing}{$\mathcal{Q}_p,\,\partial\mathcal{Q}$} \Comment{Eqs.~\eqref{eq:edge_parametric_line}--\eqref{eq:quadratic_root}, Table~\ref{tab:bc_method}, Fig.~\ref{fig:test_bc}}
    \If{$b_1\lor b_2\lor b_3$} \Comment{predicate~\eqref{eq:ccpp_predicate}}
      \State $\mathcal{W}^{\mathrm{ep}}_f\leftarrow\mathcal{W}^{\mathrm{ep}}_f\cup\{t\}$
    \EndIf
  \EndFor
  \State Group contiguous feasible epochs in $\mathcal{W}^{\mathrm{ep}}_f$ into coarse brackets $\mathcal{W}_b=\{[\tilde t^{\mathrm{in}}_m,\,\tilde t^{\mathrm{out}}_m]\}$
  \For{each bracket $[\tilde t^{\mathrm{in}}_m,\,\tilde t^{\mathrm{out}}_m]\in\mathcal{W}_b$}
    \State $t^{\mathrm{in}}_m\leftarrow$ \Call{BoundaryBisection}{$[\tilde t^{\mathrm{in}}_m-\Delta t_f,\,\tilde t^{\mathrm{in}}_m],\;exit,\;\varepsilon_t$} \Comment{Appendix~\ref{app:bisection_refinement}}
    \State $t^{\mathrm{out}}_m\leftarrow$ \Call{BoundaryBisection}{$[\tilde t^{\mathrm{out}}_m,\,\tilde t^{\mathrm{out}}_m+\Delta t_f],\;exit,\;\varepsilon_t$} \Comment{Appendix~\ref{app:bisection_refinement}}
    \State $\mathcal{W}_g\leftarrow\mathcal{W}_g\cup\{[t^{\mathrm{in}}_m,\,t^{\mathrm{out}}_m]\}$
  \EndFor
  \State $i\leftarrow i+1$
\EndWhile
\State \Return $\mathcal{W}_g$
\end{algorithmic}
\end{breakablealgorithm}

\section{Per-Window Imaging Quality Evaluation and Effective Window Screening}
\label{sec:geosot_evaluation}

The window generation procedure of Section~\ref{sec:window_generation} produces geometry feasible candidate windows, but these windows are not interchangeable for regional stripmap reconnaissance. Different observation geometries lead to markedly different focusing behavior and achievable resolution. This section develops a signal level SAR simulation that quantifies imaging performance across candidate windows and derives physically grounded screening constraints for effective window formation.

\subsection{Key geometric parameters governing imaging quality}
\label{sec:geometry_level}

A side looking stripmap acquisition mode is assumed throughout this work.  The antenna boresight is fixed relative to the platform and the ground swath is illuminated continuously as the satellite advances along its orbit.  Under this configuration, the imaging performance of each candidate window is primarily governed by four geometric parameters.  Since these quantities have already been defined in Section~\ref{sec:coarse_screening}, only their quality implications are summarized here.

The incidence angle $\theta_{\mathrm{inc}}$, related to the ground point elevation angle by Eq.~\eqref{eq:el_inc_relation}, controls the projection from slant range to ground range resolution:
\begin{equation}
\delta_r = \frac{c}{2B}, \qquad
\delta_g = \frac{\delta_r}{\sin\theta_{\mathrm{inc}}},
\label{eq:range_resolution}
\end{equation}
where $c$ is the speed of light in vacuum, $B$ is the transmitted chirp bandwidth, $\delta_r$ is the slant range resolution, $\delta_g$ is the ground range resolution, and $\theta_{\mathrm{inc}}$ is the incidence angle. Larger incidence angles lead to coarser ground range resolution for the same radar hardware. The squint angle $\theta_{\mathrm{sq}}$, related to the Doppler cone angle $\theta_{\mathrm{DC}}$ in Eq.~\eqref{eq:doppler_cone}, governs Doppler centroid behavior and range azimuth coupling in stripmap imaging. Larger squint magnitudes imply less favorable focusing conditions. The slant range $R$, defined as the Euclidean distance from the satellite position to the target ECEF position at the window midpoint, affects both propagation loss and apparent target angular motion. For a fixed platform speed, a larger slant range yields a narrower Doppler bandwidth and therefore a broader azimuth mainlobe. Finally, the dwell time, defined as the duration over which the target remains within the azimuth beam footprint, bounds the theoretical azimuth resolution through
\begin{equation}
\delta_a = \frac{L_a}{2},
\label{eq:azimuth_res}
\end{equation}
where $L_a$ is the physical antenna length, and approaching this bound requires a coherent integration time at least as long as the prescribed minimum dwell. These geometric interpretations follow standard SAR imaging principles~\cite{curlander1991synthetic,cumming2005digital}.

The geometric quantities above are evaluated at the temporal midpoint of each candidate window using the same SGP4 orbit propagator and ECEF coordinate transformations introduced in Section~\ref{sec:coarse_screening}.

\subsection{Echo Generation and Back Projection Focusing}
\label{sec:echo_bp}

The per-window quality evaluation requires a signal-level simulation chain that generates raw echo data for a point target at the regional centroid and focuses the echo through a time-domain BP algorithm. The processing steps follow established SAR signal processing methodology~\cite{li2023raw_data_sim,li2024accurate_range} and are implemented in an open-source Python package accompanying this work. The key steps are summarized below.

The raw echo is acquired in the two-dimensional fast-time and slow-time coordinate system. Fast time $\tau$ runs along each received pulse and is sampled at the analog-to-digital converter rate $F_s$. The fast-time window is centered on the two-way propagation delay to the scene center and spans $N_r$ range samples. Slow time $\eta$ indexes successive transmitted pulses; the pulse repetition interval is the reciprocal of the pulse repetition frequency, and the slow-time origin is anchored to the zero-Doppler epoch of the candidate window. The window duration determines the azimuth pulse count $N_a$. The radar sensor parameters used in the simulation are listed in Table~\ref{tab:exp3_sar_params}.

The transmitted waveform is a linear frequency modulated chirp. The baseband chirp signal and its chirp rate are defined by
\begin{equation}
s_{\mathrm{tx}}(\tau)=\exp(j\pi K_r\tau^2),\qquad
K_r=\frac{B}{\tau_p},
\label{eq:chirp_signal}
\end{equation}
where $B$ is the bandwidth and $\tau_p$ the pulse duration. Echo generation adopts a frequency-domain formulation. For each azimuth pulse the reference chirp spectrum $P(f)$ is precomputed once. For each scatterer within the instantaneous antenna beam footprint, the two-way propagation delay determines a frequency-dependent phase shift, and the per-pulse echo spectrum is assembled as
\begin{equation}
E(f)=\sum_k \sigma_k\,P(f)\,\exp\!\bigl(-j2\pi(f_0+f)\Delta t_k\bigr),
\label{eq:echo_spectrum}
\end{equation}
where $\sigma_k$ is the complex reflectivity of scatterer $k$, $f_0$ the carrier frequency, $\Delta t_k$ the two-way propagation delay of scatterer $k$, and the sum extends over all scatterers illuminated by the antenna mainlobe at that pulse. The time-domain echo is recovered by inverse fast Fourier transform. Zero-reflectivity scatterers are pruned before the pulse loop to avoid wasted computation. The slant range to each scatterer is evaluated with either a flat-Earth hyperbolic range model or a full three-dimensional Euclidean distance from the satellite ECEF position propagated by the SGP4 model at each pulse epoch, consistent with the orbit propagation framework of Section~\ref{sec:window_generation}. An optional sinc-squared antenna pattern taper can be applied in both azimuth and elevation to reproduce natural radiometric variation across the illuminated swath.

BP focusing processes the range-compressed echo through coherent time-domain summation over the synthetic aperture. Range compression applies a matched filter in the frequency domain,
\begin{equation}
H_{\mathrm{rc}}(f)=P^*(f),
\label{eq:matched_filter}
\end{equation}
followed by a factor-of-four spectral upsampling of the range profile to achieve subpixel interpolation accuracy. The output image is formed on a regular ground-plane grid at the native sensor resolution spacing. The grid is oriented with one axis along the satellite ground track and the other along the cross-track direction; pixel positions are expressed in ECEF coordinates from the SGP4-propagated satellite state. For each pulse at slow time $\eta_i$, the three-dimensional slant range $R_{i,m,n}$ from the satellite to pixel $\bm{p}_{m,n}$ is computed, the upsampled range profile is linearly interpolated at the corresponding two-way delay index, and the interpolated value receives a phase rotation for coherent accumulation:
\begin{equation}
\begin{gathered}
I(m,n)=\sum_{i=0}^{N_a-1}
s_{\mathrm{rc}}\bigl(\eta_i,\;2R_{i,m,n}/c\bigr)\,
\exp\!\Bigl(j\frac{4\pi}{\lambda}R_{i,m,n}\Bigr),\\[2pt]
R_{i,m,n}=\|\bm{r}_s(\eta_i)-\bm{p}_{m,n}\|,
\end{gathered}
\label{eq:bp_summation}
\end{equation}
where $s_{\mathrm{rc}}(\eta,\tau)$ denotes the range-compressed and upsampled echo at slow time $\eta$ and fast time $\tau$, $\bm{r}_s(\eta_i)$ is the satellite ECEF position at pulse $i$, and $\lambda$ is the carrier wavelength. The focused complex image $I(m,n)$ then supplies the point target impulse response profiles from which the quality indicators defined in Section~\ref{sec:evaluation_level} are extracted.

\subsection{Quality evaluation indicators and effective imaging time}
\label{sec:evaluation_level}

Each candidate window is assigned a companion point target for imaging quality evaluation. A unit reflectivity scatterer is placed at the geodetic center of the target region, mapped to ECEF coordinates, and evaluated with the same orbit parameters, observation geometry, echo generation model, and BP imaging process as the area scene. The azimuth pulse count~$N_a$ is obtained by multiplying the pulse repetition frequency by the candidate window duration and rounding the product up to an integer slow time extent. This extent is then bounded below by a modest minimum pulse count and above by a fixed cap. The resulting simulation remains tractable, while the retained coherent aperture still spans far more pulses than the minimum dwell required by the geometric resolution screening. The range sample count~$N_r$ is held fixed at the common fast time record length shared by the echo synthesizer for the companion point and for the regional area patch. Let~$h$ denote the focused impulse response in either the range or the azimuth direction through the peak, normalized to unit peak amplitude. The mainlobe is the central portion bounded by the first nulls, and the sidelobe region comprises the remainder of the profile. Three indicators are defined as follows.

\begin{figure*}
\centering
\begin{subfigure}[t]{0.31\textwidth}
\centering
\includegraphics[width=\linewidth]{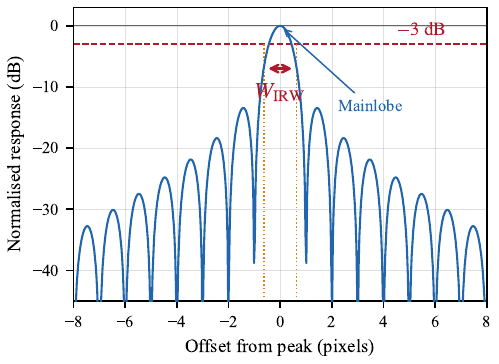}
\caption{$W_{\mathrm{IRW}}$.}
\label{fig:metric_irw}
\end{subfigure}
\hfill
\begin{subfigure}[t]{0.31\textwidth}
\centering
\includegraphics[width=\linewidth]{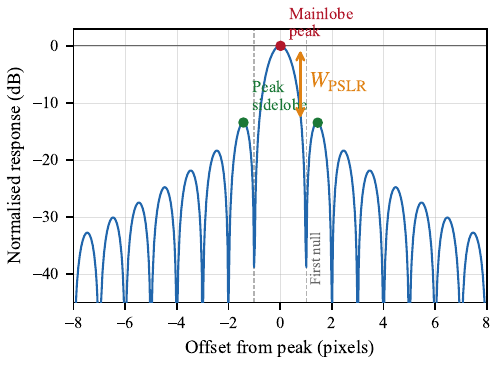}
\caption{$W_{\mathrm{PSLR}}$.}
\label{fig:metric_pslr}
\end{subfigure}
\hfill
\begin{subfigure}[t]{0.31\textwidth}
\centering
\includegraphics[width=\linewidth]{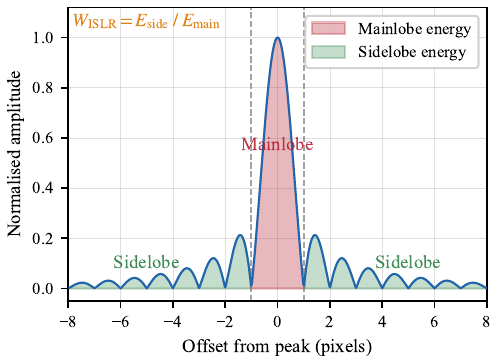}
\caption{$W_{\mathrm{ISLR}}$.}
\label{fig:metric_islr}
\end{subfigure}
\caption{Schematic definitions of the three point target quality indicators.}
\label{fig:metric_definitions}
\end{figure*}

\begin{itemize}
\item \textbf{Impulse response width.}
The impulse response width~$W_{\mathrm{IRW}}$ characterizes the effective focused resolution and is measured as the mainlobe width at the $-3$\,dB level, as shown in Fig.~\ref{fig:metric_definitions}\subref{fig:metric_irw}.
Formally,
\begin{equation}
W_{\mathrm{IRW}} = \Delta x_{-3\,\mathrm{dB}}.
\label{eq:irw}
\end{equation}

\item \textbf{Peak sidelobe ratio.}
The peak sidelobe ratio~$W_{\mathrm{PSLR}}$ characterizes the strongest sidelobe relative to the mainlobe peak and is evaluated with the mainlobe boundary set by the first nulls, as shown in Fig.~\ref{fig:metric_definitions}\subref{fig:metric_pslr}.
Formally,
\begin{equation}
W_{\mathrm{PSLR}} = \frac{\max_{\text{sidelobe}} |h|^2}{\max_{\text{mainlobe}} |h|^2}.
\label{eq:pslr_indicator}
\end{equation}

\item \textbf{Integrated sidelobe ratio.}
The integrated sidelobe ratio~$W_{\mathrm{ISLR}}$ characterizes the total sidelobe energy relative to the mainlobe energy and is computed from the integrated squared amplitude over the two regions, as shown in Fig.~\ref{fig:metric_definitions}\subref{fig:metric_islr}.
Formally,
\begin{equation}
W_{\mathrm{ISLR}} = \frac{\displaystyle\int_{\text{sidelobe}} |h|^2\,dx}{\displaystyle\int_{\text{mainlobe}} |h|^2\,dx}.
\label{eq:islr_indicator}
\end{equation}
\end{itemize}

The theoretical quantities $\delta_g$ and $\delta_a$ in Eqs.~\eqref{eq:range_resolution} and~\eqref{eq:azimuth_res} are geometric resolution bounds derived from bandwidth and antenna length arguments. By contrast, $W_{\mathrm{IRW}}$ is the realized $-3$\,dB mainlobe width measured from the focused discrete impulse response after the complete echo generation and BP processing chain. The two quantities are related, but they are not numerically identical. In particular, $W_{\mathrm{IRW}}$ also reflects aperture truncation, residual squint variation within the window, interpolation in range, and output grid sampling. Accordingly, $\delta_g$ and $\delta_a$ are used here as interpretable geometric descriptors, whereas IRW, PSLR, and ISLR serve as the final quality indicators at the signal level.

In the regional SAR scheduling problem considered here, the basic object entering the subsequent optimization stage is the candidate observation window rather than an internal sub interval of that window. The preprocessing stage assigns each window a compact and comparable quality description for screening and ranking before scheduling. The candidate window is used as the evaluation unit, and the three indicators above characterize its overall imaging quality. Effective imaging time is retained only as an auxiliary descriptor of aperture support for interpreting the resulting IRW, PSLR, and ISLR values.

For each geometry feasible candidate window $w\in\mathcal{W}_g$, the companion point target yields a six-dimensional per window quality descriptor,
\begin{equation}
\begin{gathered}
\bm{m}_w=
\bigl(
W^{\mathrm{rg}}_{\mathrm{IRW},w},\;
W^{\mathrm{az}}_{\mathrm{IRW},w},\;
W^{\mathrm{rg}}_{\mathrm{PSLR},w},\\
\qquad
W^{\mathrm{az}}_{\mathrm{PSLR},w},\;
W^{\mathrm{rg}}_{\mathrm{ISLR},w},\;
W^{\mathrm{az}}_{\mathrm{ISLR},w}
\bigr),
\end{gathered}
\end{equation}
where the superscripts $\mathrm{rg}$ and $\mathrm{az}$ denote the range and azimuth cuts, respectively.  Let
$\overline{W}^{\mathrm{rg}}_{\mathrm{IRW}}$, $\overline{W}^{\mathrm{az}}_{\mathrm{IRW}}$, $\overline{W}^{\mathrm{rg}}_{\mathrm{PSLR}}$, $\overline{W}^{\mathrm{az}}_{\mathrm{PSLR}}$, $\overline{W}^{\mathrm{rg}}_{\mathrm{ISLR}}$, and $\overline{W}^{\mathrm{az}}_{\mathrm{ISLR}}$ denote the mission-dependent acceptance thresholds.  The retained effective window set is then defined by
\begin{equation}
\mathcal{W}_{\mathrm{eff}}=
\Bigl\{
w\in\mathcal{W}_g:\;
\begin{gathered}
W^{\mathrm{rg}}_{\mathrm{IRW},w}\le \overline{W}^{\mathrm{rg}}_{\mathrm{IRW}},\\
W^{\mathrm{az}}_{\mathrm{IRW},w}\le \overline{W}^{\mathrm{az}}_{\mathrm{IRW}},\\
W^{\mathrm{rg}}_{\mathrm{PSLR},w}\le \overline{W}^{\mathrm{rg}}_{\mathrm{PSLR}},\\
W^{\mathrm{az}}_{\mathrm{PSLR},w}\le \overline{W}^{\mathrm{az}}_{\mathrm{PSLR}},\\
W^{\mathrm{rg}}_{\mathrm{ISLR},w}\le \overline{W}^{\mathrm{rg}}_{\mathrm{ISLR}},\\
W^{\mathrm{az}}_{\mathrm{ISLR},w}\le \overline{W}^{\mathrm{az}}_{\mathrm{ISLR}}
\end{gathered}
\Bigr\}.
\label{eq:window_quality_gate}
\end{equation}
where $\mathcal{W}_{\mathrm{eff}}$ denotes the retained effective window set used for subsequent scheduling. This acceptance rule links the four geometry quantities in Section~\ref{sec:geometry_level} with the six point target quality metrics above without an additional aggregation stage. It converts visible intervals directly into effective observation windows.

\section{Experiments}
\label{sec:experiments}

This section reports the experimental validation of the proposed framework. The experiments follow the two stage preprocessing chain of the framework. Experiment~I evaluates the window generation pipeline by comparing the candidate window discretization procedure with an independent reference in terms of geometry feasible window generation. Experiment~II evaluates the imaging quality assessment stage by examining geometry dependent quality metrics and the resulting effective window screening behavior. All experiments are executed on a 64 bit Windows laptop with an Intel 12th Gen Core i9-12900H CPU (base frequency 2.5~GHz, 14 physical cores, 20 logical processors). Satellite state propagation employs the SGP4 model. Candidate window generation and geometry dependent per window quality analysis are implemented in Python~3.11.

\subsection{Experiment I: Validation of Fast Interval Reduction and Geometry Feasible Window Generation}
\label{sec:exp2}

Experiment~I evaluates the proposed fast window generation pipeline against an independent reference tool. A Sentinel-1 SAR satellite is evaluated over one full day against a compact regional target. The resulting candidate observation windows are compared in number, timing, and duration with the access intervals computed by the STK side looking SAR payload module under identical geometric constraints.

\subsubsection{Experimental setup}

The inputs of Algorithm~\ref{alg:ccpp_refined_screening} are summarized in Tables~\ref{tab:exp2_satellite} and~\ref{tab:exp2_scenario}. The satellite descriptor is built from the latest publicly available TLE of Sentinel-1 downloaded from spacetrack.org, corresponding to the sun synchronous LEO platform with NORAD identifier~39634. In addition to the orbital state, the satellite input also specifies the STK consistent ground elevation and Doppler cone limits used in the side looking SAR access definition. The scenario descriptor then fixes the 24 h propagation interval, the rectangular regional target near Beijing, and the three numerical parameters governing coarse screening, refined sampling, and boundary bisection.

\begin{table*}
\centering
\caption{Experiment~I: Satellite descriptor $\mathcal{S}$.}
\label{tab:exp2_satellite}
\begin{tabular}{@{}p{0.28\linewidth}p{0.64\linewidth}@{}}
\toprule
Descriptor item & Value \\
\midrule
TLE & {\footnotesize 1 39634U 14016A 26105.96752153 -.00000200 00000-0 -32698-4 0 9995\newline 2 39634 98.1805 114.4030 0001403 85.6204 274.5156 14.59198107640942} \\
Ground elevation limits & $\theta_{\mathrm{el},\min}=15^{\circ}$, $\theta_{\mathrm{el},\max}=43^{\circ}$ \\
Doppler cone limits & $\theta_{\mathrm{DC,fwd}}=87^{\circ}$, $\theta_{\mathrm{DC,bwd}}=93^{\circ}$ \\
\bottomrule
\end{tabular}
\end{table*}

\begin{table*}
\centering
\caption{Experiment~I: Scenario descriptor $\mathcal{X}$.}
\label{tab:exp2_scenario}
\begin{tabular}{@{}p{0.36\linewidth}p{0.56\linewidth}@{}}
\toprule
Descriptor item & Value \\
\midrule
Propagation interval & [2026-04-15 00:00:00 UTC, 2026-04-16 00:00:00 UTC] \\
Regional target vertices & $T_1=(39.7^{\circ}\mathrm{N},116.2^{\circ}\mathrm{E})$; $T_2=(39.7^{\circ}\mathrm{N},116.6^{\circ}\mathrm{E})$ \\
 & $T_3=(40.1^{\circ}\mathrm{N},116.6^{\circ}\mathrm{E})$; $T_4=(40.1^{\circ}\mathrm{N},116.2^{\circ}\mathrm{E})$ \\
Angular safety margin $\epsilon_m$ & $1^{\circ}$ \\
Coarse scan step $\Delta t_c$ & 15\,s \\
Refined sampling step $\Delta t_f$ & 3\,s \\
Bisection tolerance $\varepsilon_t$ & 0.1\,s \\
\bottomrule
\end{tabular}
\end{table*}

\begin{figure*}
\centering
\includegraphics[width=0.8\textwidth]{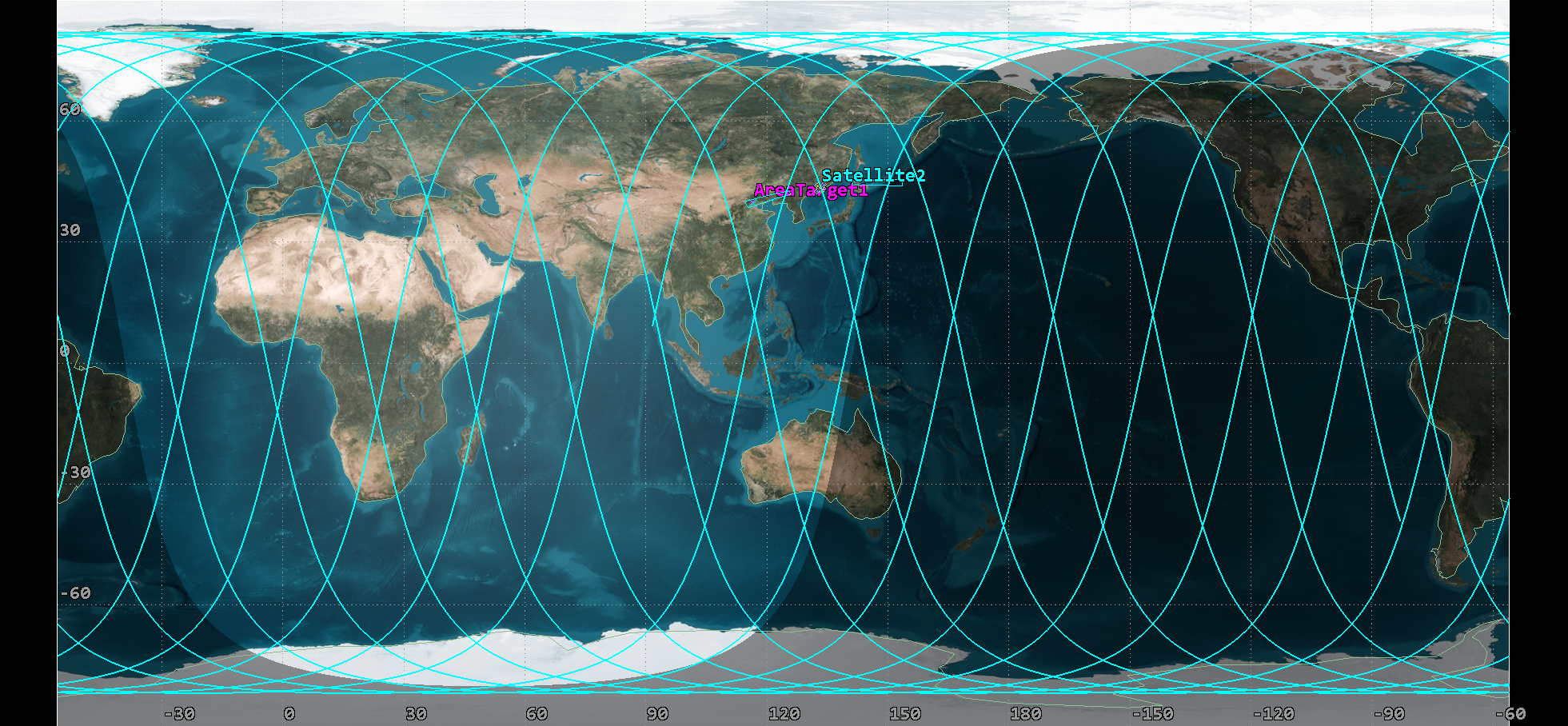}
\caption{STK reference scenario used in Experiment~I.}
\label{fig:exp2_stk_scene}
\end{figure*}

In the STK reference configuration, the regional target is created as an AreaTarget object in the Pattern mode using the four corner vertices listed in Table~\ref{tab:exp2_scenario}. The specific STK reference scene is shown in Fig.~\ref{fig:exp2_stk_scene}. To keep the access evaluation consistent with the geometric definitions of Section~\ref{sec:window_generation}, the ground station position and local geodetic up direction are interpreted under the same WGS-84 geodetic convention used in Eq.~\eqref{eq:geodetic_to_ecef}. Satellite motion is supplied to STK through an external ephemeris in .e format generated by the Python sgp4 library rather than by the native STK SGP4 propagator~\cite{ansys_stk_ephemeris_file}, because preliminary comparison identified a measurable mismatch in the latter.

\subsubsection{Results and Stepwise Analysis of the Three Stage Method}

Both the proposed pipeline and the STK reference produce exactly three access windows over the 24 h period. Table~\ref{tab:exp2_comparison} lists the start epoch, stop epoch, and duration of each window from both methods. The final column reports the absolute timing difference.

\begin{table}
\centering
\caption{Experiment~I: Window comparison between the proposed method and the STK reference.}
\label{tab:exp2_comparison}
\footnotesize
\setlength{\tabcolsep}{4pt}
\begin{tabular}{@{}c l l l r r@{}}
\toprule
\# & Source & Start (UTC) & Stop (UTC) & Duration & $\Delta t$ \\
\midrule
\multirow{2}{*}{1}
  & STK    & 08:36:21.228 & 08:36:55.594 & 34.366\,s &  \\
  & Proposed & 08:36:21.234 & 08:36:55.594 & 34.359\,s & 6\,ms \\
\midrule
\multirow{2}{*}{2}
  & STK    & 22:12:48.969 & 22:13:10.833 & 21.864\,s &  \\
  & Proposed & 22:12:48.961 & 22:13:10.828 & 21.867\,s & 8\,ms \\
\midrule
\multirow{2}{*}{3}
  & STK    & 23:50:14.127 & 23:50:47.261 & 33.133\,s &  \\
  & Proposed & 23:50:14.133 & 23:50:47.262 & 33.129\,s & 6\,ms \\
\bottomrule
\end{tabular}
\end{table}

\begin{figure*}
\centering
\begin{subfigure}[t]{0.9\textwidth}
\centering
\includegraphics[width=0.95\linewidth]{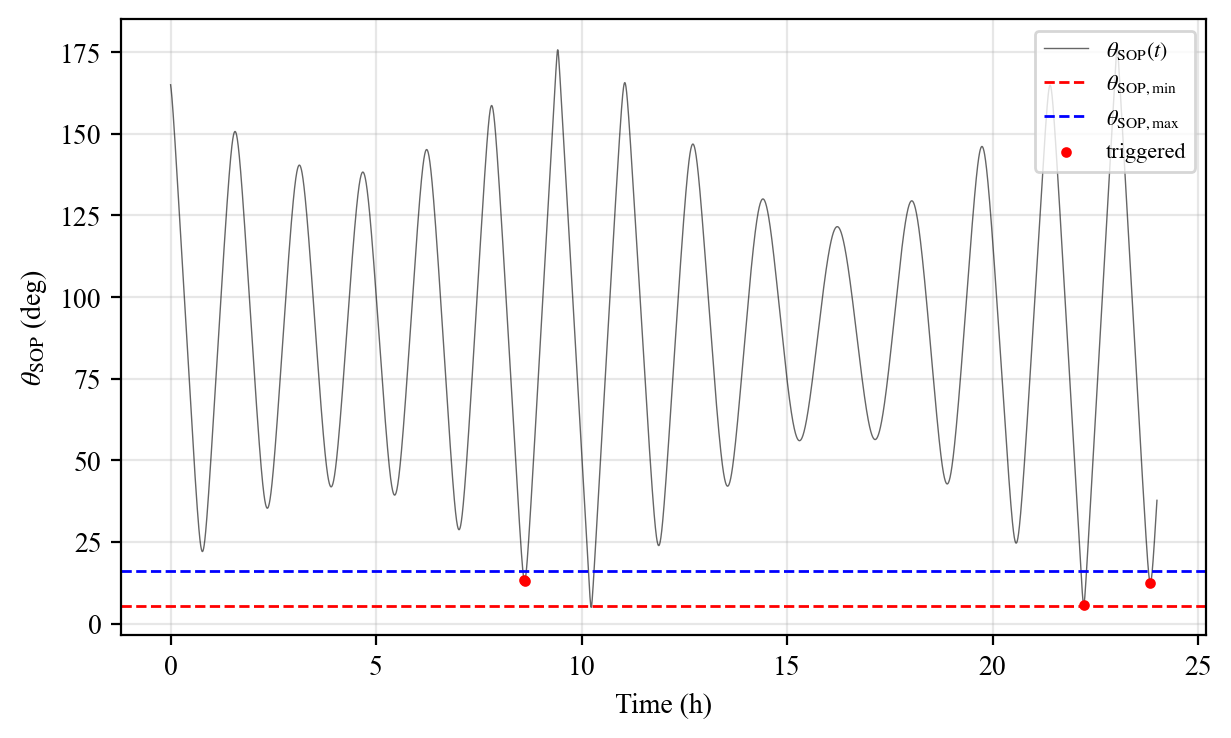}
\caption{Cross track angle $\theta_{\mathrm{SOP}}$.}
\label{fig:exp2_jsvf_sop}
\end{subfigure}

\vspace{0.3em}

\begin{subfigure}[t]{0.9\textwidth}
\centering
\includegraphics[width=0.95\linewidth]{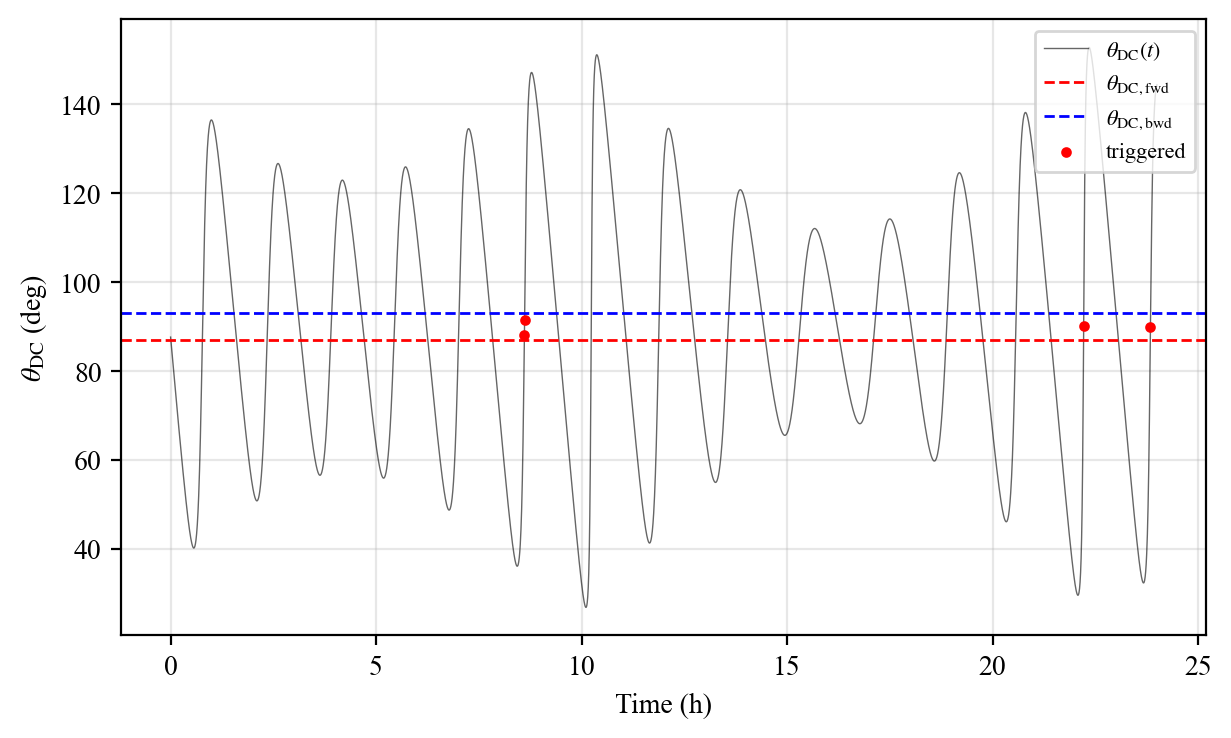}
\caption{Doppler cone angle $\theta_{\mathrm{DC}}$.}
\label{fig:exp2_jsvf_dc}
\end{subfigure}
\caption{Time histories used in the coarse angular screening.}
\label{fig:exp2_jsvf_timeseries}
\end{figure*}

\begin{figure*}
\centering
\begin{subfigure}[t]{0.48\textwidth}
\centering
\includegraphics[width=\linewidth]{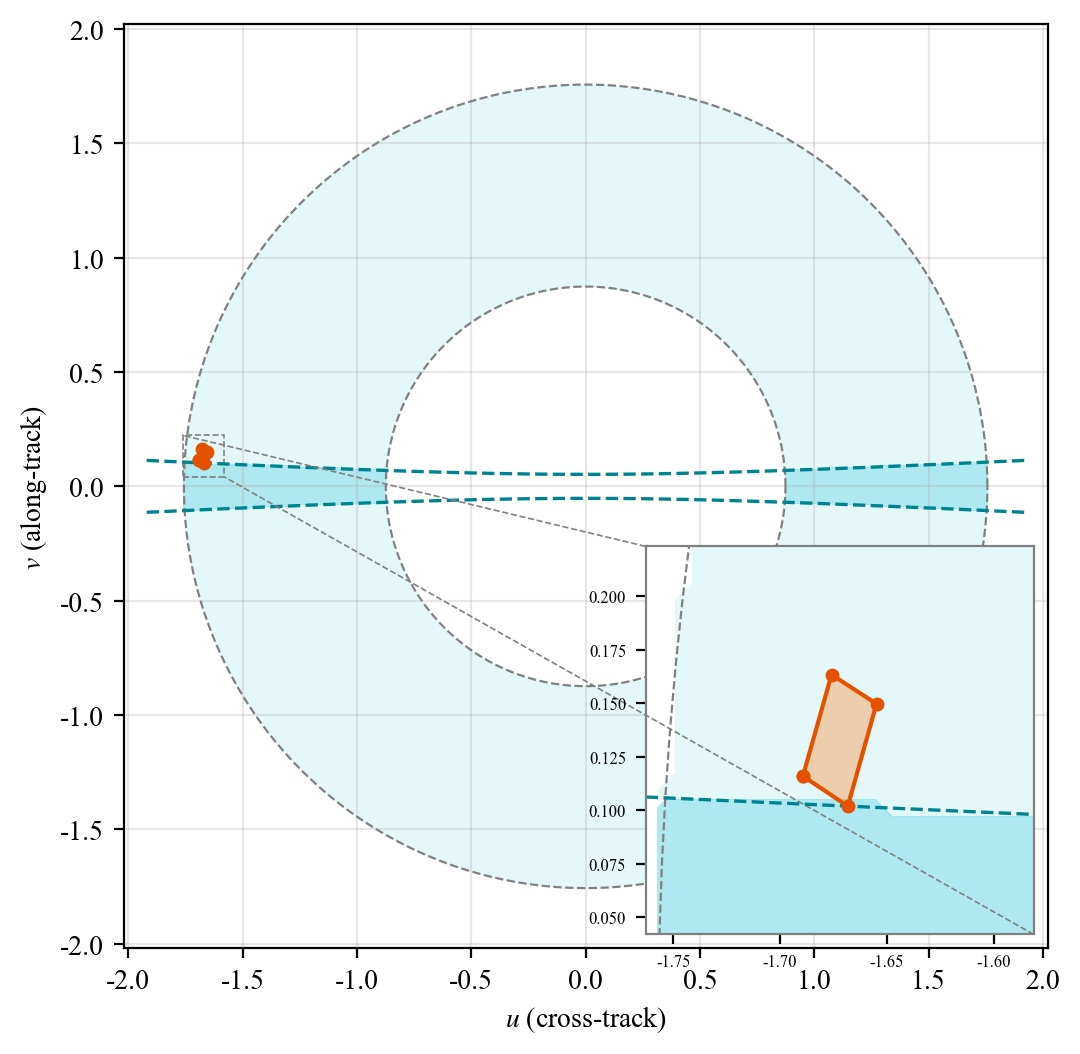}
\caption{Entry instant of Window~1.}
\end{subfigure}
\hfill
\begin{subfigure}[t]{0.48\textwidth}
\centering
\includegraphics[width=\linewidth]{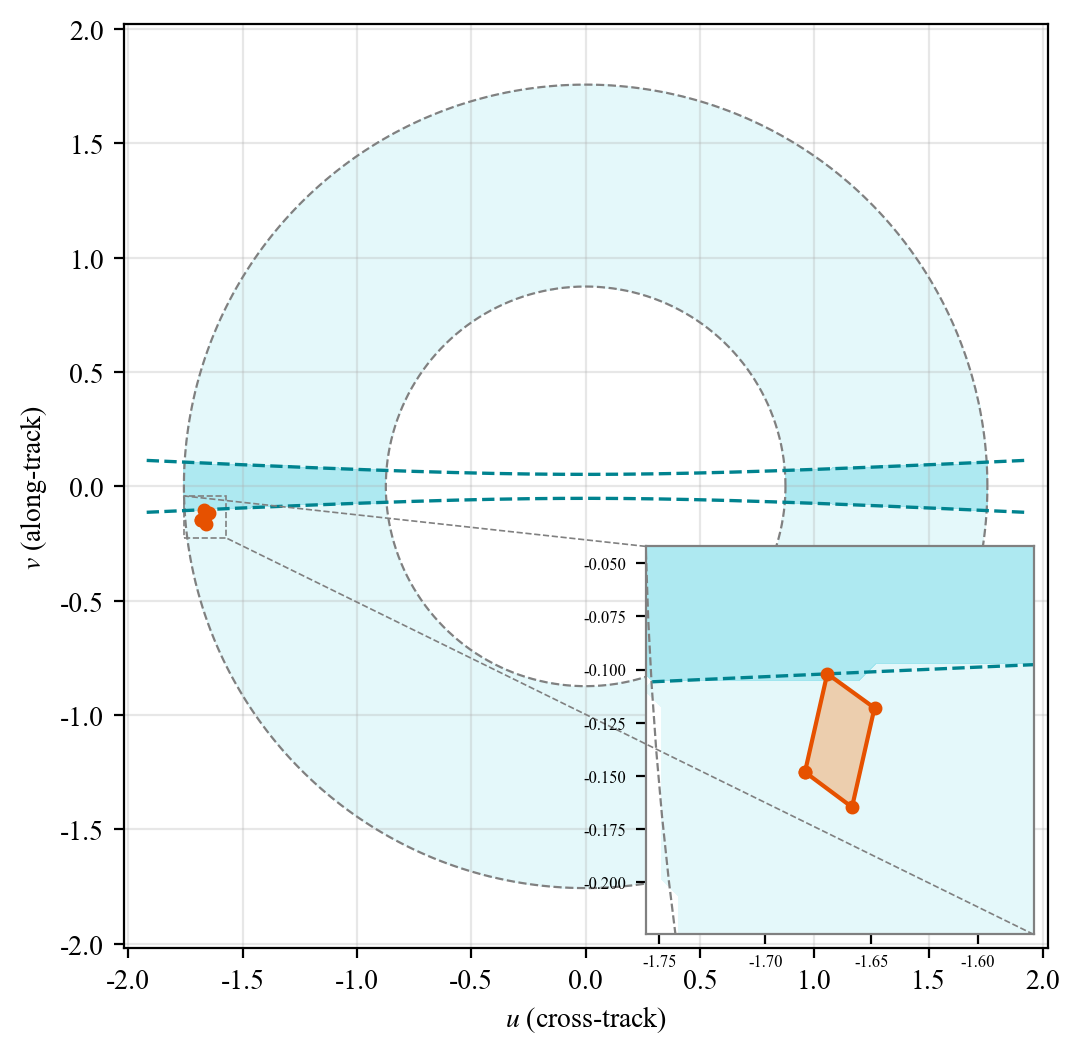}
\caption{Exit instant of Window~1.}
\end{subfigure}
\caption{VVLH calculation plane projections at the entry and exit instants of Window~1.}
\label{fig:exp2_fov}
\end{figure*}

\begin{figure*}
\centering
\includegraphics[width=0.7\textwidth]{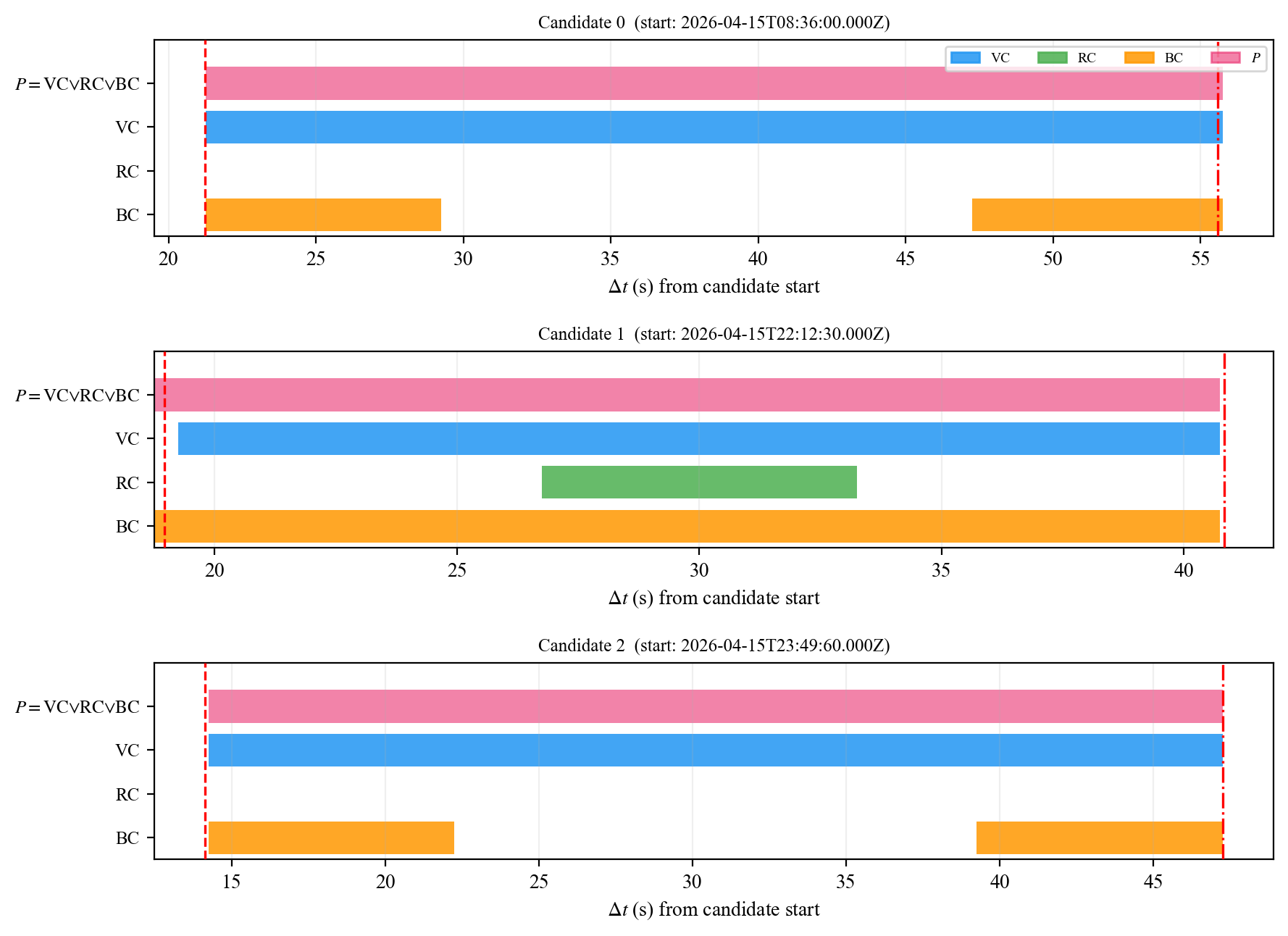}
\caption{Refined predicate decomposition over two candidate intervals.}
\label{fig:exp2_predicate}
\end{figure*}

The two methods agree on the number and temporal ordering of all windows. The largest start time discrepancy is 8\,ms and the largest end time discrepancy is 5\,ms, both arising at Window~2. These sub centisecond differences are consistent with the bisection tolerance of 0.1\,s and are well below the typical orbit determination uncertainty for LEO satellites. The duration differences are bounded by 7\,ms. Figs.~\ref{fig:exp2_jsvf_sop}--\ref{fig:exp2_predicate} and Table~\ref{tab:exp2_bisection} detail the three stage reduction.

Step~1 is the coarse angular screening. Figs.~\ref{fig:exp2_jsvf_sop} and~\ref{fig:exp2_jsvf_dc} show the 24 hour time histories of $\theta_{\mathrm{SOP}}$ and $\theta_{\mathrm{DC}}$, where the dashed lines denote the corresponding bandpass bounds and the red dots mark epochs that satisfy both coarse conditions simultaneously. The joint filter retains four coarse candidate windows over the full day and removes the remaining infeasible orbit arcs before any polygon level computation is invoked.

Step~2 is the refined predicate evaluation on the VVLH calculation plane. Fig.~\ref{fig:exp2_fov} shows the entry and exit instants of Window~1, with the admissible domain rendered as the shaded region, the projected target polygon as the orange quadrilateral, and the magnified boundary crossing displayed in the inset of each panel. In both panels, the admissible domain is bounded by the inner and outer circles ($R_{\mathrm{in}}$, $R_{\mathrm{out}}$) and clipped by the forward and backward Doppler hyperbolas, while the orange quadrilateral is the projected regional target. At the refined entry epoch, the polygon first penetrates the admissible domain; at the refined exit epoch, the last boundary contact disappears. These two boundary events jointly determine the geometry feasible window returned by the refined predicate.

Fig.~\ref{fig:exp2_predicate} decomposes this second stage into three constituent tests, namely vertex containment (VC), reverse containment (RC), and boundary crossing (BC), whose logical union yields the full predicate (P); the dashed lines in the figure indicate the refined window boundaries returned by bisection. Vertex containment spans the largest portion of each candidate interval because at least one projected vertex enters the admissible set over most feasible epochs. Boundary crossing extends the true interval by several seconds near the temporal margins and recovers the exact entry and exit neighbourhoods. Reverse containment is activated only over a short interior segment of Candidate~1, where the projected target polygon momentarily encloses a representative point of the admissible domain. The VC trace in the same figure shows that at least one projected vertex already lies inside the admissible set over that segment, so VC and RC are simultaneously true there; by contrast, BC is no longer the dominant contributor on that interior portion because it is away from the entry and exit boundaries. In other words, for this example RC acts as a concurrent confirmation of the feasible interior segment rather than as the sole mechanism recovering an interval missed by both VC and BC. RC is still retained because more general geometric configurations can produce genuine reverse containment cases that are detected only by RC.

\begin{table*}
\centering
\caption{Experiment~I: Bisection bracketing pairs and refined window boundaries.}
\label{tab:exp2_bisection}
\footnotesize
\setlength{\tabcolsep}{4pt}
\begin{tabular}{@{}c l l l l c@{}}
\toprule
Window & Entry bracket (UTC) & Refined entry & Exit bracket (UTC) & Refined exit & Iterations \\
\midrule
1 & [08:36:21, 08:36:24] & 08:36:21.234 & [08:36:54, 08:36:57] & 08:36:55.594 & 10 \\
2 & [22:12:48, 22:12:51] & 22:12:48.961 & [22:13:09, 22:13:12] & 22:13:10.828 & 10 \\
3 & [23:50:12, 23:50:15] & 23:50:14.133 & [23:50:45, 23:50:48] & 23:50:47.262 & 10 \\
\bottomrule
\end{tabular}
\end{table*}

Step~3 is the bisection refinement of the window boundaries. Table~\ref{tab:exp2_bisection} lists the initial entry and exit bracketing pairs generated from the 3 s refined sampling grid and the final boundaries returned by bisection. With a stopping tolerance of 0.1~s, each window requires 10 iterations in total for its entry and exit boundaries. The refined windows obtained from these time pairs are exactly those reported in Table~\ref{tab:exp2_comparison}, with millisecond level residual differences relative to the STK reference.

Experiment~I recovers every STK access interval with end point errors of order $10$\,ms and with no missed or spurious window. Elevation and Doppler cone filters eliminate most orbit arcs before any polygon intersection is evaluated, the geometric predicates are then applied only to the short retained intervals, and bisection isolates the crossings where feasibility changes. Under the same SAR geometry assumptions, the resulting procedure matches the reference while preserving an auditable preprocessing stage for the subsequent imaging quality screening and effective window screening.

\subsection{Experiment II: Signal Level Imaging Quality Validation Across Three Observation Windows}
\label{sec:exp3}

Experiment~II validates the imaging quality differentiation method described in Section~\ref{sec:geosot_evaluation} by executing a complete echo generation and focusing cycle on all three geometry feasible windows produced by Experiment~I for the Beijing AOI. Different observation geometries, although all satisfying the sensor visibility envelope, produce measurably different imaging quality outcomes.

\subsubsection{Time Varying Observation Geometry}
\label{sec:exp3_geom}

For each of the three geometry feasible windows produced by Experiment~I, four governing geometric descriptors introduced in Section~\ref{sec:geometry_level} are sampled at 60 uniformly spaced instants within the window. These descriptors are the incidence angle $\theta_{\mathrm{inc}}$, the squint angle $\theta_{\mathrm{sq}}$, the slant range $R$ from the satellite to the regional centroid, and the ground range resolution $\delta_g$ inferred from Eq.~\eqref{eq:range_resolution}. The satellite state at each sampling instant is obtained from the SGP4 plus GMST pipeline used throughout Section~\ref{sec:window_generation}, and the descriptors are evaluated between that state and the ECEF position of the regional centroid. This yields four time series per window that together characterise how the observation geometry evolves during each candidate dwell, and provide the basis for a compact midpoint based summary in Table~\ref{tab:exp3_window_summary}.

\begin{figure*}
\centering
\begin{subfigure}[b]{0.4\textwidth}
  \centering
  \includegraphics[width=\linewidth]{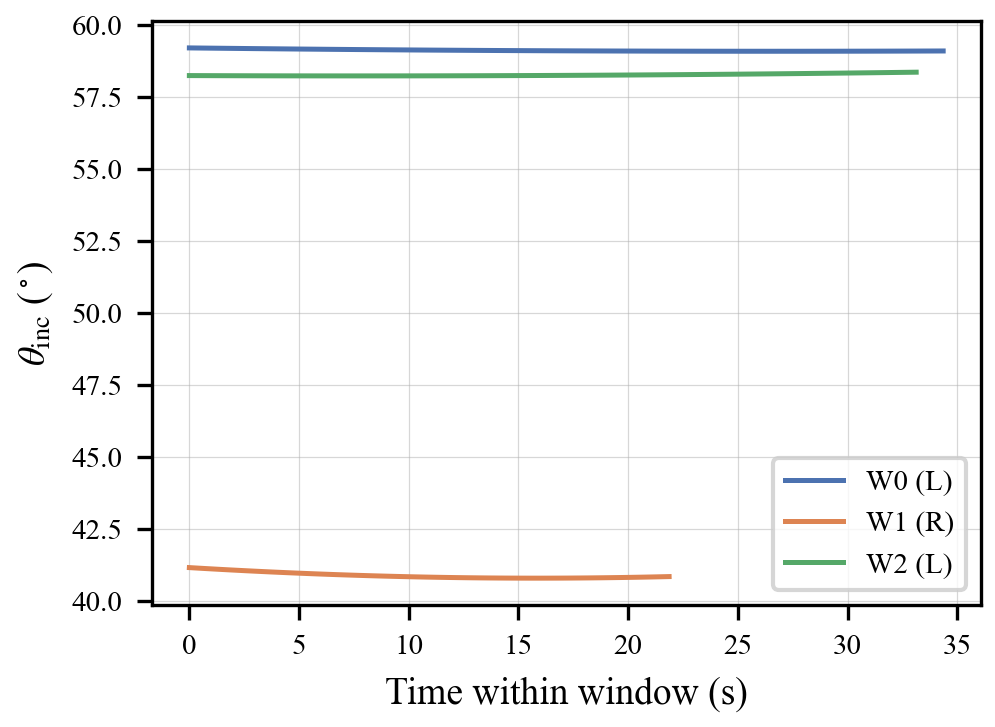}
  \caption{$\theta_{\mathrm{inc}}$}
  \label{fig:geom_inc}
\end{subfigure}\hfill
\begin{subfigure}[b]{0.4\textwidth}
  \centering
  \includegraphics[width=\linewidth]{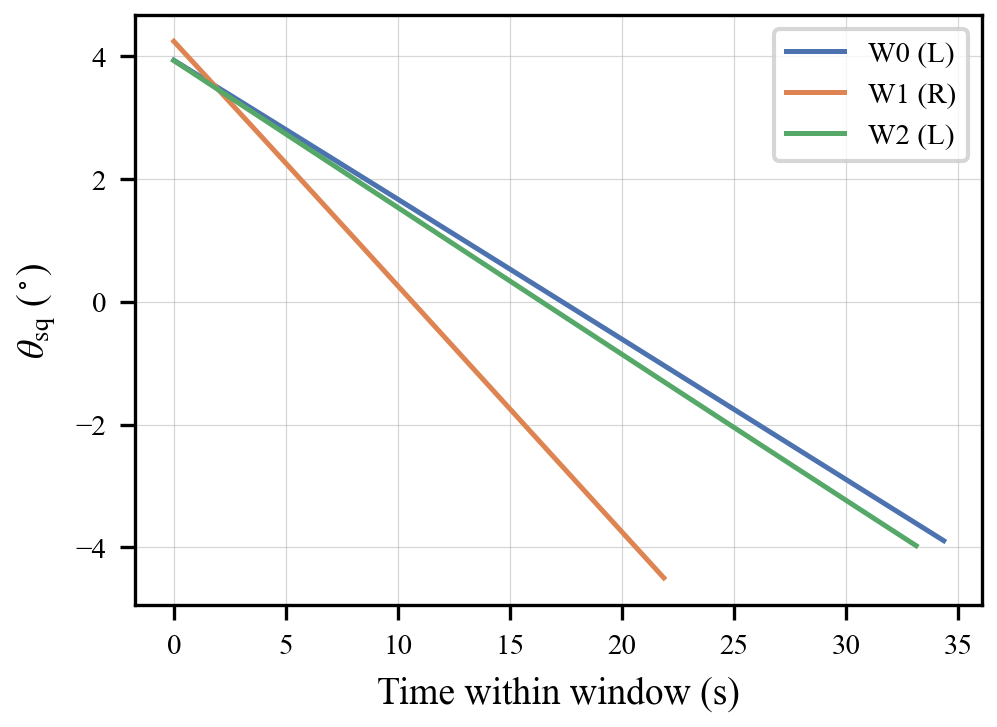}
  \caption{$\theta_{\mathrm{sq}}$}
  \label{fig:geom_sq}
\end{subfigure}

\vspace{4pt}

\begin{subfigure}[b]{0.4\textwidth}
  \centering
  \includegraphics[width=\linewidth]{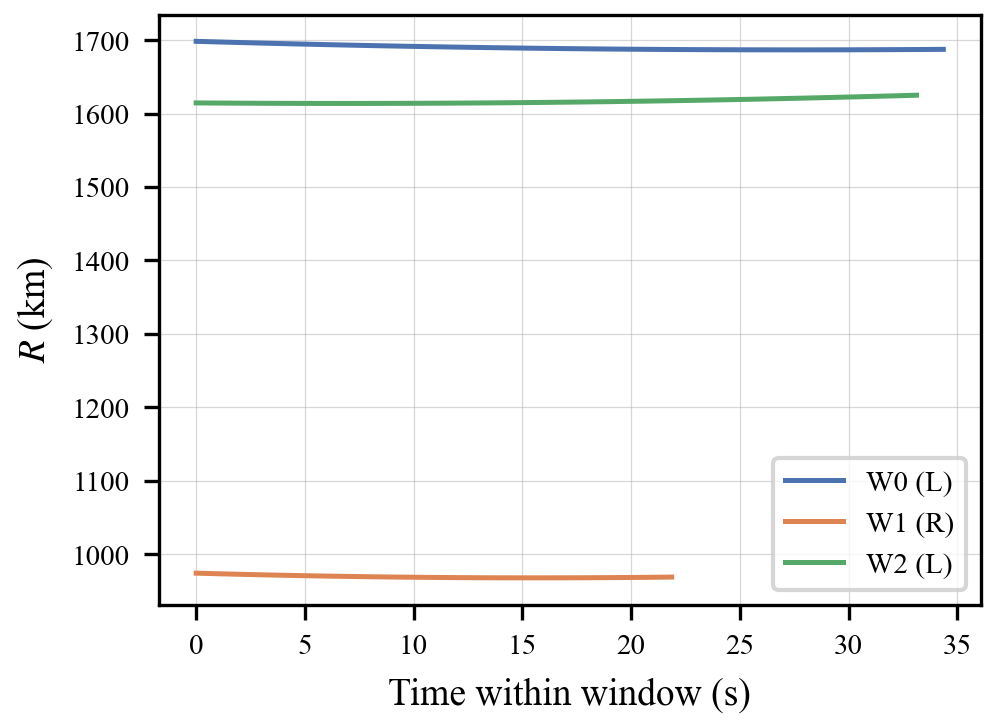}
  \caption{$R$}
  \label{fig:geom_R}
\end{subfigure}\hfill
\begin{subfigure}[b]{0.4\textwidth}
  \centering
  \includegraphics[width=\linewidth]{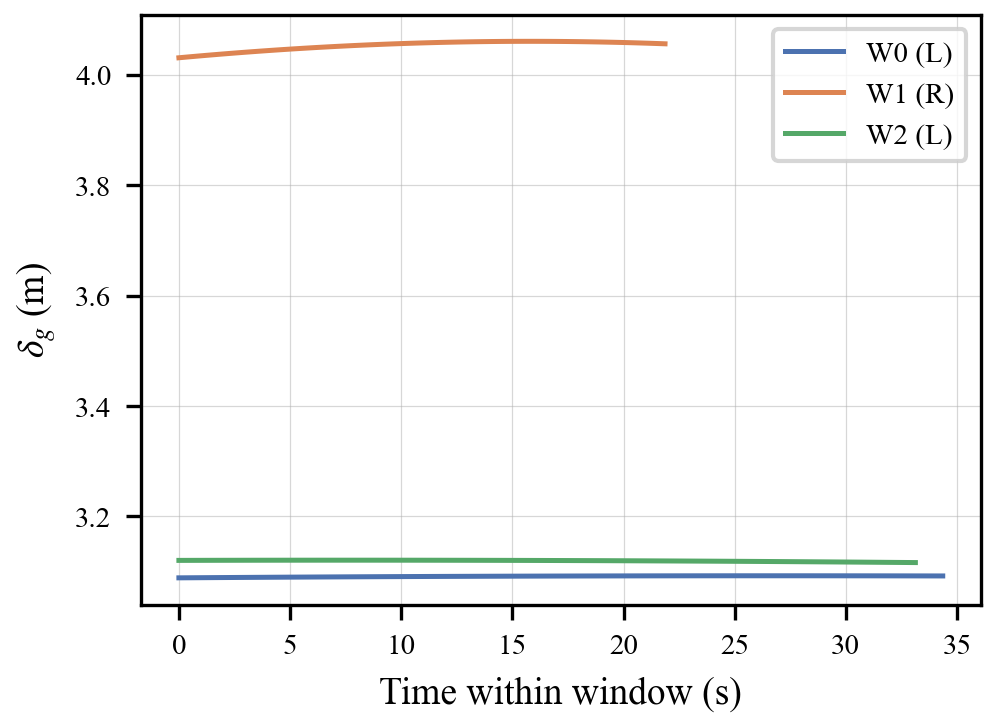}
  \caption{$\delta_g$}
  \label{fig:geom_dg}
\end{subfigure}
\caption{Time varying observation geometry within the three candidate windows.}
\label{fig:geom_timeseries}
\end{figure*}

The resulting time series are plotted in Figs.~\ref{fig:geom_inc}--\ref{fig:geom_dg}. Over the 20 to 34\,s window durations, the incidence angle and slant range vary by less than $0.4^{\circ}$ and 5\,km, respectively, so both quantities stay effectively constant within a window. The ground range resolution also remains nearly invariant within each window, and Table~\ref{tab:exp3_window_summary} reports midpoint values for inter window comparison. The inter window contrast remains substantial: W1 at a moderate incidence of $40.8^{\circ}$ yields $4.06$\,m, whereas W0 and W2 at approximately $59^{\circ}$ yield roughly $3.1$\,m. The squint angle sweeps approximately linearly from $+4^{\circ}$ to $-4^{\circ}$, crossing zero Doppler near the window center and reflecting the azimuth synthesis geometry intrinsic to stripmap SAR integration.

The window descriptors summarized later in Table~\ref{tab:exp3_window_summary} retain the midpoint quantities most relevant to inter window comparison. Windows W0 and W2 are left looking passes at large incidence angles of 59.1 and 58.2\,degrees, while W1 is a right looking pass at a moderate incidence of 40.8\,degrees.

\subsubsection{Echo generation and focusing}

The SAR sensor parameters adopted in the simulation are listed separately in Table~\ref{tab:exp3_sar_params}, aligned with the Sentinel-1 C band stripmap mode.

\begin{table}
\centering
\caption{SAR simulation parameters adopted in Experiment~II, aligned with Sentinel-1 stripmap mode.}
\label{tab:exp3_sar_params}
\begin{tabular}{lll}
\toprule
Parameter & Symbol & Value \\
\midrule
Carrier frequency & $f_0$ & 5.405\,GHz \\
Chirp bandwidth & $B$ & 56.5\,MHz \\
Pulse duration & $\tau_p$ & 41.75\,$\mu$s \\
Pulse repetition frequency & PRF & 1720\,Hz \\
Antenna length & $L_a$ & 12.3\,m \\
\bottomrule
\end{tabular}
\end{table}

\begin{table}
\centering
\caption{Quality screening parameters used in Experiment~II.}
\label{tab:quality_screening_parameters}
\footnotesize
\begin{tabular}{@{}p{0.32\columnwidth}p{0.34\columnwidth}p{0.20\columnwidth}@{}}
\toprule
Threshold symbol & Meaning & Value \\
\midrule
$\overline{W}^{\mathrm{rg}}_{\mathrm{IRW}}$ & Range IRW threshold & $5.5$\,m \\
$\overline{W}^{\mathrm{az}}_{\mathrm{IRW}}$ & Azimuth IRW threshold & $9.0$\,m \\
$\overline{W}^{\mathrm{rg}}_{\mathrm{PSLR}}$ & Range PSLR threshold & $-13.0$\,dB \\
$\overline{W}^{\mathrm{az}}_{\mathrm{PSLR}}$ & Azimuth PSLR threshold & $-11.0$\,dB \\
$\overline{W}^{\mathrm{rg}}_{\mathrm{ISLR}}$ & Range ISLR threshold & $-10.0$\,dB \\
$\overline{W}^{\mathrm{az}}_{\mathrm{ISLR}}$ & Azimuth ISLR threshold & $-10.0$\,dB \\
\bottomrule
\end{tabular}
\end{table}

\begin{figure}
\centering
\includegraphics[width=0.40\textwidth]{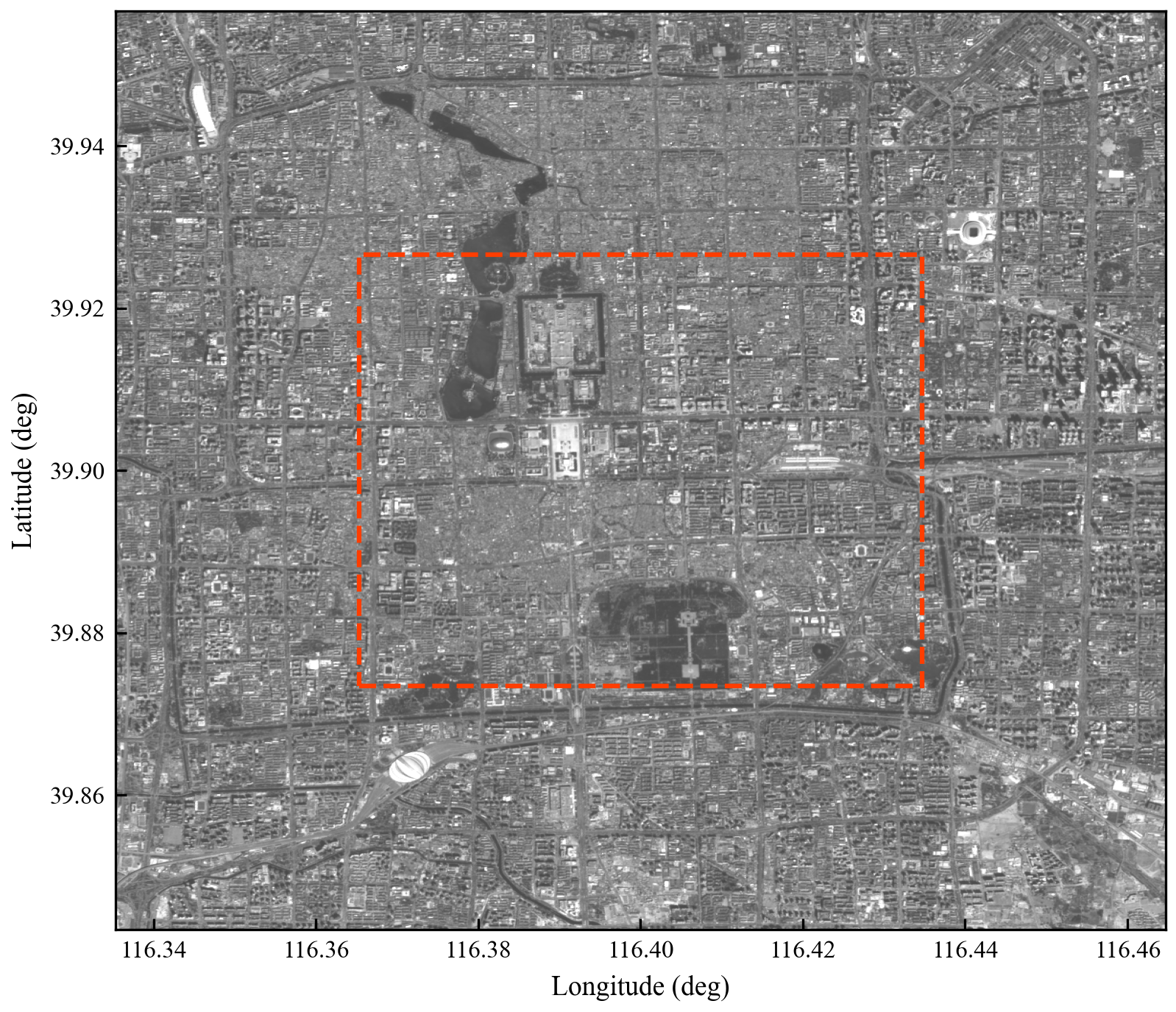}
\caption{Sentinel-2C L2A optical basemap over central Beijing.}
\label{fig:s2_optical}
\end{figure}

Table~\ref{tab:quality_screening_parameters} lists the six acceptance thresholds used for the effective window filter in Experiment~II. These thresholds correspond directly to Eq.~\eqref{eq:window_quality_gate} and are stated in the same physical units as the point target metrics reported later in Table~\ref{tab:exp3_window_summary}. The scene reflectivity is taken from a Sentinel-2C MSI Level 2A product at 10\,m resolution.  Fig.~\ref{fig:s2_optical} shows the grayscale optical basemap over central Beijing, where the dashed rectangle marks the simulation patch used for area target echo generation.  The scene is represented by a regular grid of point scatterers at 5\,m spacing over an approximately 8.8\,km patch centered on the AOI, each weighted by the corresponding Sentinel-2 surface reflectance.  An additional unit reflectivity point target is placed at the scene center for quantitative quality extraction.  Echoes are generated with a frequency domain per scatterer model using the radar parameters in Table~\ref{tab:exp3_sar_params}, and BP focusing is performed on a 400 by 400 output grid at 15\,m pixel spacing.

\begin{table*}
\centering
\caption{Experiment~II summary of midpoint window descriptors and point target quality metrics for the three candidate windows over the Beijing AOI.}
\label{tab:exp3_window_summary}
\footnotesize
\setlength{\tabcolsep}{5pt}
\begin{tabular}{llccc}
\toprule
Category & Parameter & W0 & W1 & W2 \\
\midrule
\multirow{5}{*}{Window descriptor}
 & UTC window & 08:36:21--08:36:56 & 22:12:49--22:13:11 & 23:50:14--23:50:47 \\
 & Incidence angle (deg) & 59.1 & 40.8 & 58.2 \\
 & Look side & Left & Right & Left \\
 & $\delta_g$ (m) & 3.09 & 4.06 & 3.12 \\
 & $\delta_a$ (m) & 6.15 & 6.15 & 6.15 \\
\midrule
\multirow{6}{*}{Point target quality}
 & Range IRW (m) & 5.25 & 4.75 & 5.22 \\
 & Azimuth IRW (m) & 8.87 & 8.96 & 8.87 \\
 & Range PSLR (dB) & $-$21.5 & $-$13.1 & $-$21.1 \\
 & Azimuth PSLR (dB) & $-$11.5 & $-$12.1 & $-$11.5 \\
 & Range ISLR (dB) & $-$15.2 & $-$10.2 & $-$14.9 \\
 & Azimuth ISLR (dB) & $-$10.5 & $-$10.9 & $-$10.6 \\
\bottomrule
\end{tabular}
\end{table*}

\begin{figure*}
\centering
\begin{subfigure}[b]{0.26\textwidth}
  \includegraphics[width=\textwidth]{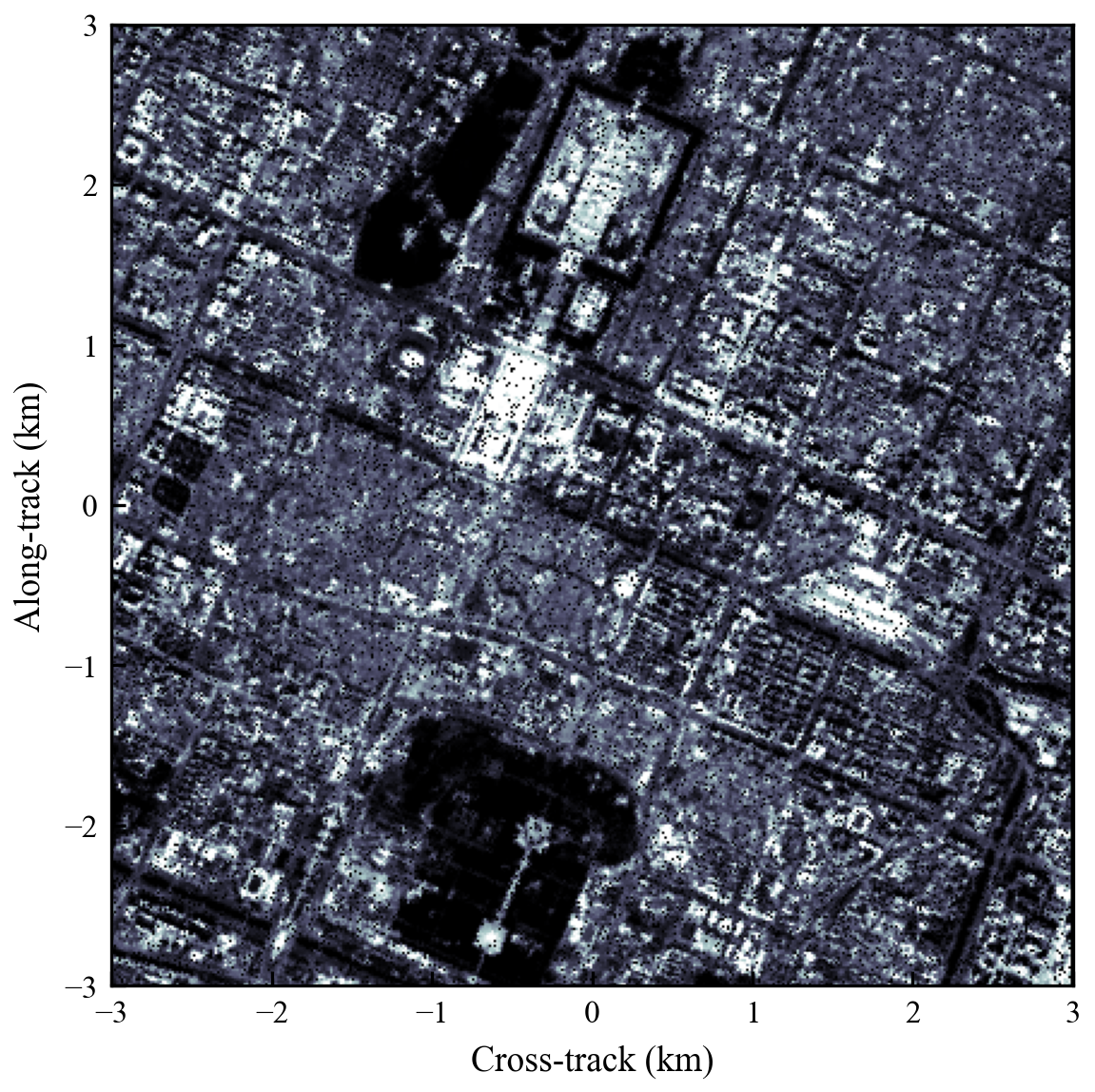}
  \caption{W0 reprojected truth, left, 59.1\,deg}
\end{subfigure}\hfill
\begin{subfigure}[b]{0.26\textwidth}
  \includegraphics[width=\textwidth]{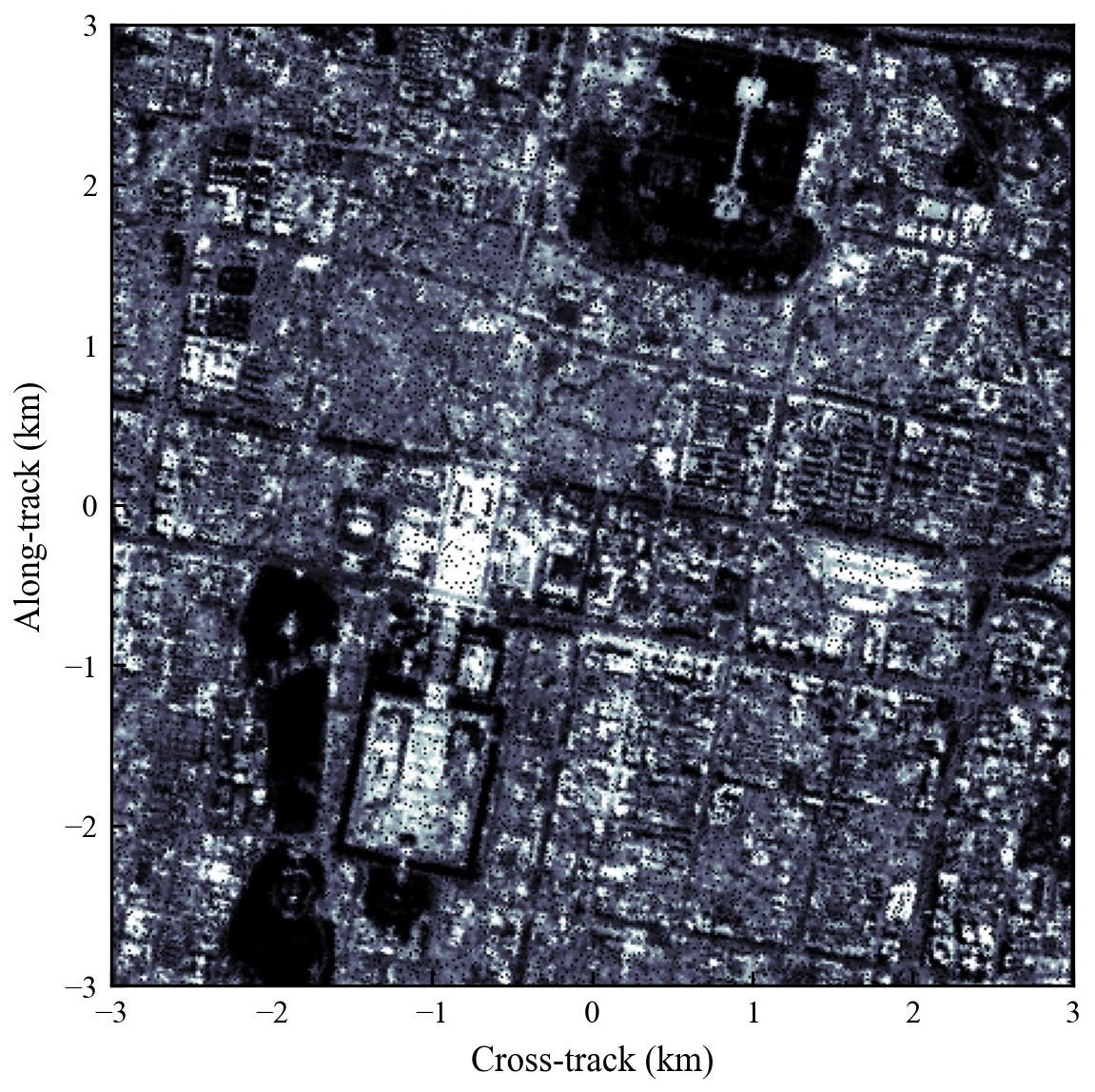}
  \caption{W1 reprojected truth, right, 40.8\,deg}
\end{subfigure}\hfill
\begin{subfigure}[b]{0.26\textwidth}
  \includegraphics[width=\textwidth]{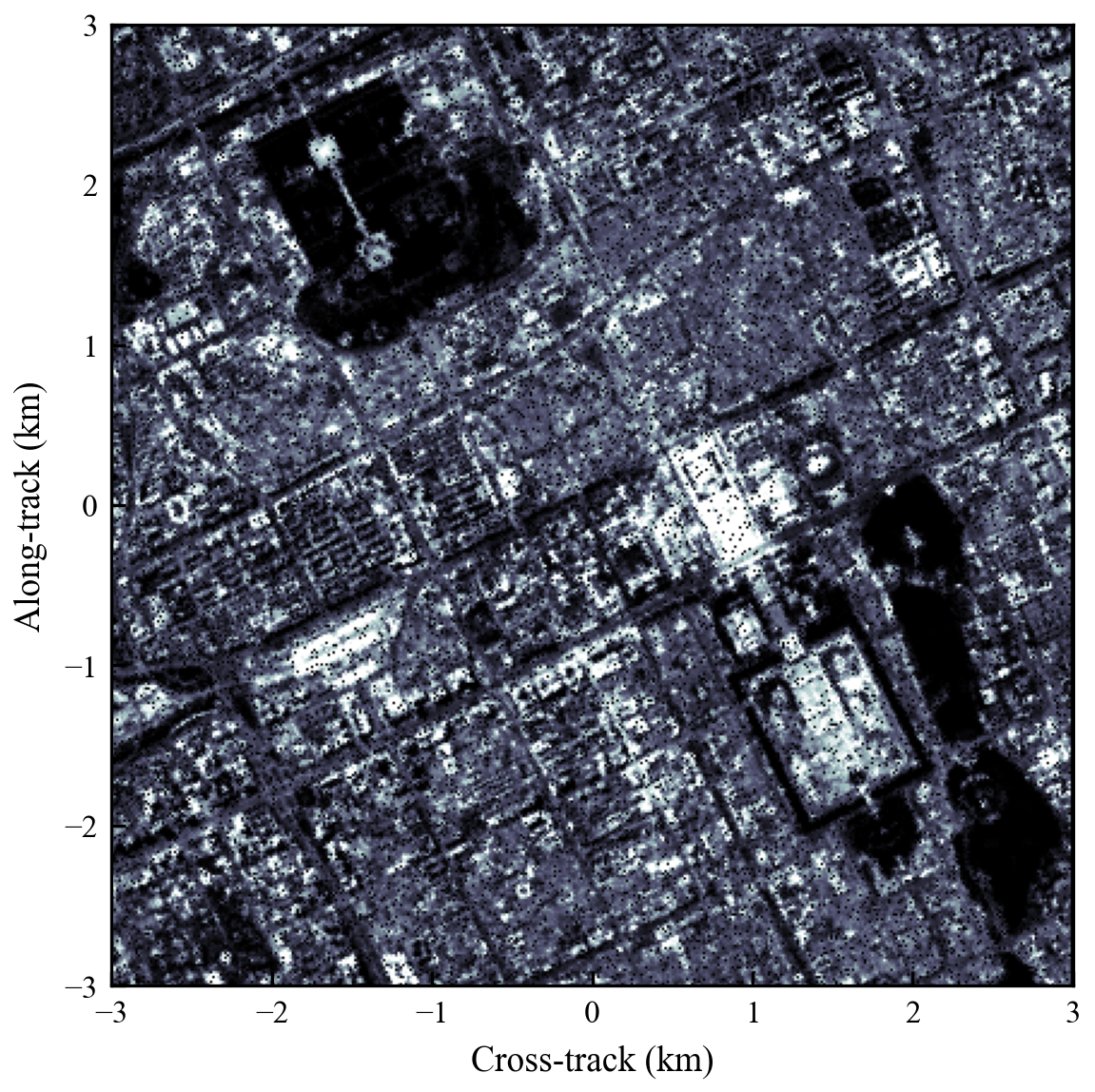}
  \caption{W2 reprojected truth, left, 58.2\,deg}
\end{subfigure}

\vspace{0.3em}

\begin{subfigure}[b]{0.26\textwidth}
  \includegraphics[width=\textwidth]{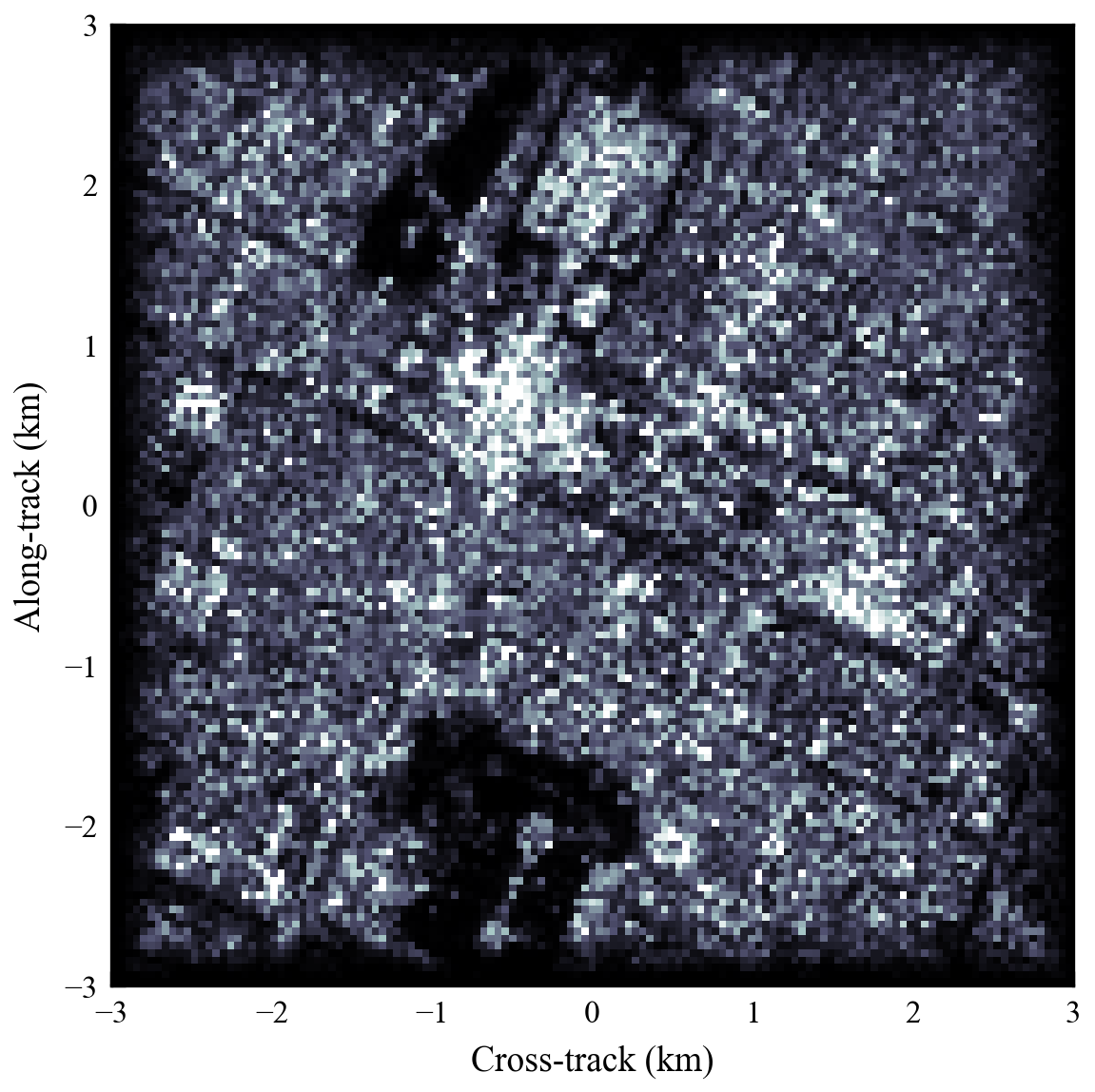}
  \caption{W0 multi look BP, left, 59.1\,deg}
\end{subfigure}\hfill
\begin{subfigure}[b]{0.26\textwidth}
  \includegraphics[width=\textwidth]{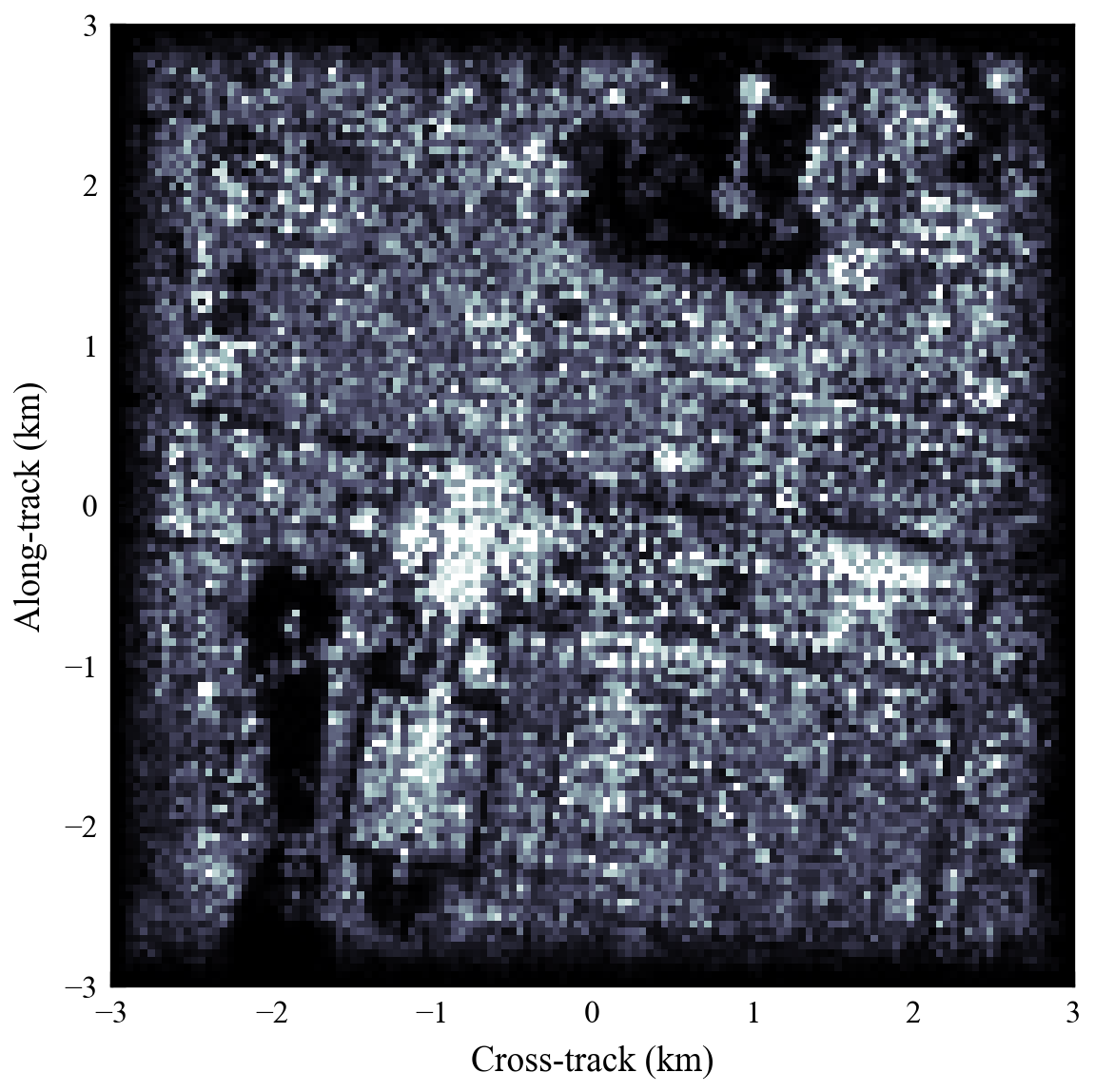}
  \caption{W1 multi look BP, right, 40.8\,deg}
\end{subfigure}\hfill
\begin{subfigure}[b]{0.26\textwidth}
  \includegraphics[width=\textwidth]{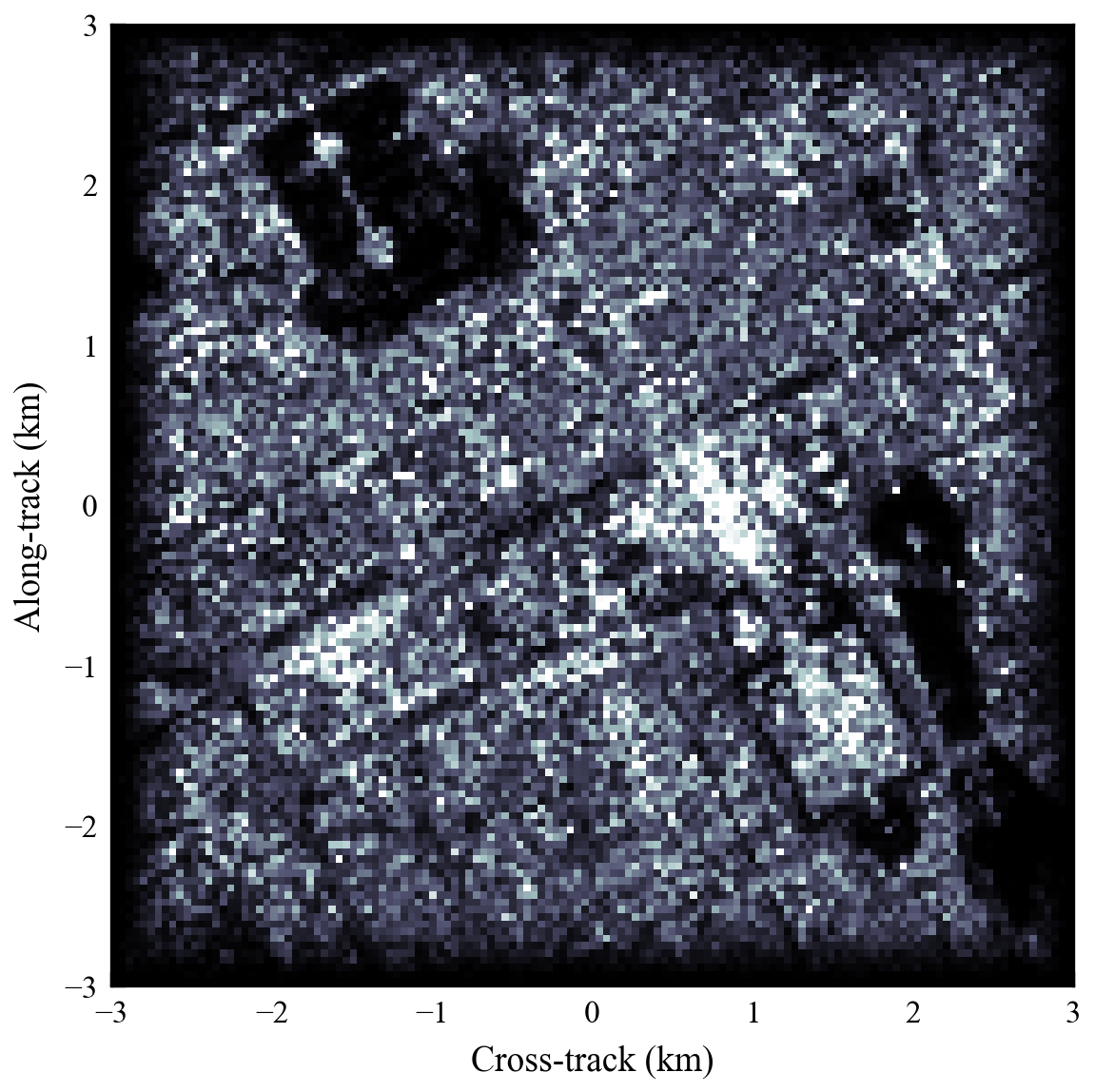}
  \caption{W2 multi look BP, left, 58.2\,deg}
\end{subfigure}
\caption{Reprojected truth maps and multi look BP images for W0 to W2.}
\label{fig:area_bp_ml}
\end{figure*}

\subsubsection{Quality evaluation results}

The point target quality rows in Table~\ref{tab:exp3_window_summary} report IRW in physical units after conversion from the native BP sampling grid. Azimuth performance is almost unchanged across the three windows. The azimuth IRW stays within a narrow range. The azimuth PSLR also remains nearly constant. The narrow spread reflects stable coherent aperture formation under the common antenna length. The main difference appears in range. W1 provides the narrowest range IRW, but it also shows substantially worse range PSLR than W0 and W2. A notable discrepancy exists between the theoretical ground range resolution and the measured range IRW. For W0 and W2, the theoretical value is approximately 3.1 m, yet the measured range IRW exceeds 5.2 m, corresponding to a broadening factor of about 1.7. The theoretical ground range resolution is a geometric bandwidth limit derived under ideal matched filtering and continuous sampling. The BP focusing chain introduces additional broadening from several sources. These include finite aperture truncation of the synthetic array, coarse output grid sampling at 15 m pixel spacing relative to a mainlobe width of approximately 3 m, range interpolation error, and residual squint variation within the integration window. These effects collectively widen the realized mainlobe well beyond the bandwidth limited prediction. The broadening is also geometry dependent. W0 and W2 operate at large incidence angles, where the ground range projection is more sensitive to slant range broadening and the grid discretization effect is more pronounced. W1 operates at a moderate incidence angle, so these projection and sampling effects are milder. This explains why W0 and W2 exhibit larger absolute broadening factors despite having finer theoretical resolution. The midpoint geometric resolution descriptor captures only the first order trend. The measured impulse response still depends on the realized sidelobe structure after finite aperture integration and discrete BP focusing. IRW, PSLR, and ISLR require joint interpretation rather than reliance on a single theoretical resolution measure.

The Sentinel-2 reflectivity scene is resampled onto the same along track and cross track grid as the SAR products by using standard slant range and azimuth geometry conventions~\cite{curlander1991synthetic,cumming2005digital}. In Fig.~\ref{fig:area_bp_ml}, the reprojected truth panels are optical reflectivity maps in the SAR image coordinates. The multi look BP panels are the corresponding focused SAR images after BP and multi looking. Fig.~\ref{fig:area_bp_ml} compares the scenes in a common image coordinate system. Under this common coordinate system, W0 and W2 at large incidence exhibit clearer road contrast, block separation, and building edge definition, while W1 at moderate incidence shows noticeably degraded sharpness and elevated sidelobe artifacts, consistent with its substantially worse range PSLR and ISLR. The visual ordering agrees with the point target metrics in Table~\ref{tab:exp3_window_summary}, where W1 approaches the acceptance thresholds in both range PSLR and range ISLR. This agreement supports the threshold based window screening rule introduced in Section~\ref{sec:geosot_evaluation}.

\section{Conclusion}
\label{sec:conclusion}

This paper presents an effective window framework for regional SAR reconnaissance that generates geometry feasible observation opportunities and evaluates per window signal level imaging quality. The framework constructs candidate windows through coarse angular screening, a planar characteristic curve containment test, and boundary bisection refinement, then evaluates each window through a companion point target echo generation and backprojection focusing pipeline that extracts spatial resolution, PSLR, and ISLR indicators. A unified effective window quality gate converts the four governing geometric descriptors and the six imaging metrics into a single acceptance decision per window.

The experiments validate the two preprocessing stages independently. Experiment~I demonstrates that the window generation pipeline matches the STK side looking SAR access reference with millisecond level accuracy while preserving an auditable geometric construction. Experiment~II confirms that the quality evaluation stage differentiates imaging performance across candidate windows and screens out opportunities that satisfy the sensor visibility envelope but produce degraded spatial resolution.

Future work will extend the window quality model to more general heterogeneous observation scenarios. One direction is a unified effective window framework for multiple payload types, incorporating illumination conditions, cloud sensitivity, viewing geometry, spatial resolution, and diverse imaging modes. Another direction is dynamic target tracking under evolving geometric constraints, revisit demand, and multi payload coordination.

\appendix
\section{Reverse Containment Ray Casting Procedure}
\label{app:reverse_containment}

For completeness, Algorithm~\ref{alg:reverse_containment_appendix} gives the detailed horizontal half ray procedure used by the ReverseContainment test in Algorithm~\ref{alg:ccpp_refined_screening}.  The representative point set is fixed deterministically as $\mathcal{Q}_0=\{(R_{\mathrm{in}},0)\}$ for connected admissible domains and as $\mathcal{Q}_0=\{(-R_{\mathrm{in}},0),(R_{\mathrm{in}},0)\}$ for the disjoint lobe configuration.

\begin{breakablealgorithm}
\caption{ReverseContainment by odd even ray casting}
\label{alg:reverse_containment_appendix}
\begin{algorithmic}[1]
\Require Projected polygon vertices $\{\bm{p}_k\}_{k=0}^{N-1}$ with $\bm{p}_k=(u_k,v_k)$ and $\bm{p}_N\equiv\bm{p}_0$; representative point set $\mathcal{Q}_0=\{\bm{q}_{0,j}\}_{j=1}^{N_q}$, chosen as $\{(R_{\mathrm{in}},0)\}$ for connected domains or $\{(-R_{\mathrm{in}},0),(R_{\mathrm{in}},0)\}$ for the disjoint lobe case
\Ensure Reverse containment indicator $b_2\in\{0,1\}$
\State $b_2\leftarrow 0$
\For{each representative point $\bm{q}_{0,j}\in\mathcal{Q}_0$}
  \State $N_{\mathrm{cross}}\leftarrow 0$
  \For{$k=0$ \textbf{to} $N-1$}
    \If{$(v_k>v_{0,j}\land v_{k+1}\leq v_{0,j})\lor(v_k\leq v_{0,j}\land v_{k+1}>v_{0,j})$} \Comment{half open crossing test}
      \State $u_{\mathrm{cross}}\leftarrow u_k + (v_{0,j}-v_k)(u_{k+1}-u_k)/(v_{k+1}-v_k)$
      \If{$u_{\mathrm{cross}}>u_{0,j}$}
        \State $N_{\mathrm{cross}}\leftarrow N_{\mathrm{cross}}+1$
      \EndIf
    \EndIf
  \EndFor
  \If{$N_{\mathrm{cross}}\bmod 2 = 1$}
    \State $b_2\leftarrow 1$
    \State \Return $b_2$
  \EndIf
\EndFor
\State \Return $b_2$
\end{algorithmic}
\end{breakablealgorithm}

\section{Boundary Bisection Refinement Procedure}
\label{app:bisection_refinement}

For completeness, Algorithm~\ref{alg:boundary_bisection_appendix} gives the detailed one dimensional bisection procedure used to refine the entry and exit boundaries of each coarse bracket in Algorithm~\ref{alg:ccpp_refined_screening}.

\begin{breakablealgorithm}
\caption{Boundary bisection for one bracket endpoint}
\label{alg:boundary_bisection_appendix}
\begin{algorithmic}[1]
\Require Initial bracket $[t_L,t_R]$; boundary type $\tau\in\{exit,exit\}$; tolerance $\varepsilon_t$
\Ensure Refined boundary epoch $t^{\star}$
\While{$t_R-t_L>\varepsilon_t$}
  \State $t_M\leftarrow (t_L+t_R)/2$
  \If{$\tau=exit$}
    \If{predicate~\eqref{eq:ccpp_predicate} is true at $t_M$}
      \State $t_R\leftarrow t_M$
    \Else
      \State $t_L\leftarrow t_M$
    \EndIf
  \Else
    \If{predicate~\eqref{eq:ccpp_predicate} is true at $t_M$}
      \State $t_L\leftarrow t_M$
    \Else
      \State $t_R\leftarrow t_M$
    \EndIf
  \EndIf
\EndWhile
\State $t^{\star}\leftarrow (t_L+t_R)/2$
\State \Return $t^{\star}$
\end{algorithmic}
\end{breakablealgorithm}

\printcredits

\section*{Declaration of competing interest}

The authors declare that they have no known competing financial interests or personal relationships that could have appeared to influence the work reported in this paper.

\section*{Acknowledgments}

This work was supported in part by the National Natural Science Foundation of China under Grant Nos. 12072366 and U2441205, and granted by State Key Laboratory of Space System Operation and Control.

\section*{Declaration of Generative AI and AI-assisted technologies in the writing process}

During the preparation of this work the authors used DeepSeek-V4 in order to improve readability and language. After using this tool, the authors reviewed and edited the content as needed and take full responsibility for the content of the publication.

\section*{Data availability}

Data and code supporting this work are available from the corresponding author upon reasonable request.

\bibliographystyle{elsarticle-num}
\bibliography{reference}

\end{document}